\documentclass[aps,prd,twocolumn,amsmath,amssymb,floatfix,nofootinbib]{revtex4}

\usepackage{graphicx}
\usepackage{dcolumn}
\usepackage{bm}
\usepackage{color}
\usepackage[export]{adjustbox}

\newcommand{\be}{\begin{eqnarray}}

\newcommand{\ee}{\end{eqnarray}}
\def\bal#1\eal{\begin{align}#1\end{align}}

\newcommand{\bsk}{\boldsymbol{k}}

\newcommand{\bea}{\begin{eqnarray}}
\newcommand{\eea}{\end{eqnarray}}
\newcommand{\barr}{\begin{array}}
\newcommand{\earr}{\end{array}}
\def\bal#1\eal{\begin{align}#1\end{align}}
\def\CL{{\cal L}}

\def\mpl{M_{\rm P}}

\newcommand{\beq}{\begin{equation}}
\newcommand{\eeq}{\end{equation}}

\def\bk{{\bf k}}

\def\CL{{\cal L}}
\def\CO{{\cal O}}

\def\mpl{M_{\rm P}}

\begin{document}

\title{Prospects for Cosmological Collider Physics}
\author{P. Daniel Meerburg$^{1}$}
\author{Moritz M{\"u}nchmeyer$^{2,3,4}$}
\author{Julian B. Mu{\~n}oz$^{5}$}
\author{Xingang Chen$^{6}$}
\affiliation{$^1$CITA, University of Toronto, 60 St.~George Street, Toronto, Canada}
\affiliation{$^2$Sorbonne Universit\'{e}s, UPMC Univ Paris 06, UMR7095}
\affiliation{$^3$CNRS, UMR7095, Institut d'Astrophysique de Paris, F-75014, Paris, France}
\affiliation{$^4$Perimeter Institute for Theoretical Physics, Waterloo ON N2L 2Y5, Canada}
\affiliation{$^5$Department of Physics and Astronomy, Johns Hopkins University, 3400 N. Charles St., Baltimore, MD 21218, USA}
\affiliation{$^6$Institute for Theory and Computation, Harvard-Smithsonian Center for Astrophysics, 60 Garden Street, Cambridge, MA 02138, USA}

\begin{abstract}
It is generally expected that heavy fields are present during inflation, which can leave their imprint in late-time cosmological observables. The main signature of these fields is a small amount of distinctly shaped non-Gaussianity, which if detected, would provide a wealth of information about the particle spectrum of the inflationary Universe.
Here we investigate to what extent these signatures can be detected or constrained using futuristic 21-cm surveys. We construct model-independent templates that extract the squeezed-limit behavior of the bispectrum, and examine their overlap with standard inflationary shapes and secondary non-Gaussianities. We then use these templates to forecast detection thresholds for different masses and couplings using a 3D reconstruction of modes during the dark ages ($z\sim 30-100$). We consider interactions of several broad classes of models and quantify their detectability as a function of the baseline of a dark ages interferometer. Our analysis shows that there exists the tantalizing possibility of discovering new particles with different masses and interactions with future 21-cm surveys.
\end{abstract}

\maketitle

\tableofcontents

\section{Introduction}

In realistic inflation models, besides one effectively-massless scalar field driving inflation, there is a vast landscape of heavy fields. Although classically these heavy fields are not very important (except when excited by sharp features in the model), their quantum fluctuations leave distinctive signatures in the density perturbations. These fluctuations are most appreciable when the masses of the fields are of the same order as the Hubble scale $H$ of the inflationary background or less.
For this reason, this class of models is called Quasi-single-field (QSF) inflation models \cite{Chen:2009we,Chen:2009zp,Baumann:2011nk,Assassi:2012zq,Chen:2012ge,Noumi:2012vr, Arkani-Hamed:2015bza,Sefusatti:2012ye,Norena:2012yi,Gong:2013sma,Emami:2013lma, Kehagias:2015jha,Dimastrogiovanni:2015pla,Schmidt:2015xka,Bonga:2015urq}. Fields with these masses may already be present in the spectrum of a UV-completed unification theory, such as the KK spectrum and stringy states in string theory \cite{Arkani-Hamed:2015bza}; they may arise from fields that are originally light, but their mass gets uplifted by loop corrections in the inflationary background \cite{Burgess:2009bs,Chen:2016nrs} or by coupling to the background curvature. The presence of supersymmetry can also provide a natural mechanism to stabilize scalar fields with mass of order $H$ \cite{Baumann:2011nk}. In addition, it has been suggested that particles with mass somewhat heavier than $\CO(H)$ may be used as the ``primordial standard clocks" \cite{Chen:2011zf,Chen:2014cwa,Chen:2015lza}  to track the evolution of the scale factor $a(t)$ of any time-dependent background, providing direct evidence for either inflation or alternative scenarios.
As such, these heavy fields could provide a wealth of information about fundamental physics and our primordial Universe.

Interestingly, heavy fields imprint potentially observable distinctive signatures in the primordial non-Gaussianities \cite{Chen:2009we,Chen:2009zp,Baumann:2011nk,Assassi:2012zq,Noumi:2012vr,Arkani-Hamed:2015bza} that are not captured in the low-energy effective theories of single-field inflation models.
In particular, Arkani-Hamed and Maldacena derived a general result \cite{Arkani-Hamed:2015bza} showing that the entire particle spectrum, including the mass and spin, is reflected in the scaling behavior of various soft limits of primordial non-Gaussianities.
Moreover, the spin of particles coupled to the inflaton can be geometrically disentangled with 3D surveys, by studying the trispectrum \cite{Jeong:2012df,Dai:2013ikl,Dai:2013kra,Dimastrogiovanni:2015pla}.
It is likely that these are the highest mass scales directly observable\footnote{In models that temporarily break scale-invariance, it is also possible that much heavier fields can be excited non-adiabatically and leave different signatures in the density perturbations \cite{Chen:2011zf,Chen:2014cwa,Flauger:2016idt}.
} in nature and provides a strong motivation to fully explore the observability of such signatures in the data. In this sense inflation has been referred to as a ``cosmological collider experiment".

In this paper we evaluate what constraints can be placed on the presence of heavy particles during inflation, considering future cosmological surveys of the 21-cm field.
In some cases the classical oscillations of the massive fields are excited by sharp features, leading to resonant signals that oscillate as a function of scale, and manifest themselves both in the power spectrum and in higher-order correlation functions. The observational constraints that will be reached on these type of features in the future have been studied in Refs.~\cite{Chen:2016zuu,Chen:2016vvw,Ballardini:2016hpi,Xu:2016kwz}. In cases where the inflation model does not contain sharp features -the main interest of this paper- the signatures of the quantum oscillations of heavy fields are more subtle.
These signals do not break scale-invariance, as quantum fluctuations themselves do not have a preferred scale.
Instead, heavy fields show their presence through shape-dependent non-Gaussianities, appearing as non-analytic scalings in various soft limits of the momentum configurations \cite{Chen:2009we,Chen:2009zp,Baumann:2011nk,Noumi:2012vr,Arkani-Hamed:2015bza}.
For example, one can infer the mass of the heavy field from the power of the momentum ratio between the long and short mode of the three-point function (or bispectrum).
For light fields in the mass spectrum this power is a real number, with the power-laws between those of the local- and equilateral-type non-Gaussianities, behaving as ``intermediate non-Gaussianities".
For heavier fields in the spectrum, however, the power becomes a complex number, which results in oscillatory signals in the bispectrum,
with the ``clock signal" being the oscillatory component of this bispectrum.
Additionally, non-zero spins generate an additional dependence on the angle between the two modes, creating an incredibly rich phenomenology \cite{Arkani-Hamed:2015bza,Lee:2016vti}.

Because heavy fields are expected in any realistic inflationary theory, and these fields must couple to the inflaton at least through gravity, the existence of the signals we study can be considered generic. However, the amplitude of these signatures depends sensitively on what type of non-linear coupling sources the non-Gaussianities, as well as on the interactions between the massive fields and the inflaton \cite{Chen:2009zp,Baumann:2011nk,Lee:2016vti}, introducing strong model dependence. The aim of this paper is to study the observational sensitivity of future experiments to the aforementioned cosmological signatures and to explore what types of couplings we will be able to constrain.

The cosmic microwave background (CMB) is currently the most powerful tool to search for primordial non-Gaussianities, and so far it is consistent with a perfectly-Gaussian Universe \cite{PlanckNGs2013,PlanckNGs2015}, at least for a large class of possible bispectra. These constraints will be extended to those including the CMB $B$-modes sourced by tensor non-Gaussianities \cite{Lee:2016vti,Meerburg:2016ecv} and a more exhaustive search will be performed to include recently-developed models (e.g.\cite{DriftingOscillations2014,Flauger:2016idt}). In the long run, both galaxy surveys \cite{LSSnonGaussianity2014}, and 21-cm cosmology \cite{21cmFNL} appear to be the most powerful tool to measure primordial non-Gaussianities.
This last probe, on which our analysis is based, consists on mapping out the distribution of neutral Hydrogen in the redshift range $30\leq z\leq 100$. During these times, most scales are linear, which should in principle contain a very predictable signal (unlike LSS at low $z$).
This provides us with an enormous amount of observable volume, which would drastically reduce cosmic variance, making it at least tempting to ask if such a futuristic cosmological probe can distinguish the interesting phenomenology occurring during inflation.

This paper is organized as follows. In Sec.~\ref{sec:review} we review the cosmological collider bispectrum results obtained in Refs.~\cite{Arkani-Hamed:2015bza,Chen:2015lza}. We then discuss overlap with well-known shapes in the literature. In Sec.~\ref{sec:template} we propose various templates that can be used to search for signatures of heavy fields in the data. In Sec.~\ref{sec:21-cm} we perform a Fisher forecast for a 21-cm experiment during the dark ages, including non-linear effects induced by gravity and the non-linear tracing of the 21-cm field to the underlying fluctuations in the density and velocity fields.
We discuss which types and sizes of interactions in inflation models are in principle observable.
We conclude in Sec.~\ref{sec:conclusions}.

\section{Non-Gaussianity from the cosmological collider}\label{sec:review}

In this section we review different inflation models with massive scalar fields and describe various perturbative diagrams that have been considered in literature. We then evaluate the shape function numerically and discuss its properties. We emphasize model-independent properties that enables us to do forecast for the observability of this general class of models. Specific details can be found in Appendix \ref{sec:analyticalshape}.

\subsection{Quasi-single-field inflation and bispectra}

The observable of interest in this paper is the curvature bispectrum from QSF inflation models, in which there are additional scalar fields with masses of order the Hubble parameter, as first proposed in Ref.~\cite{Chen:2009we}.
This setup has been considered by a number of authors \cite{Chen:2009zp,Baumann:2011nk,Assassi:2012zq,Chen:2012ge,Noumi:2012vr,Arkani-Hamed:2015bza,Sefusatti:2012ye,Norena:2012yi,Gong:2013sma,Emami:2013lma,Dimastrogiovanni:2015pla,Schmidt:2015xka,Bonga:2015urq} and we briefly review their results, with an emphasis on the characteristics relevant for observation.

\begin{figure}
\resizebox{1\hsize}{!}{
\includegraphics{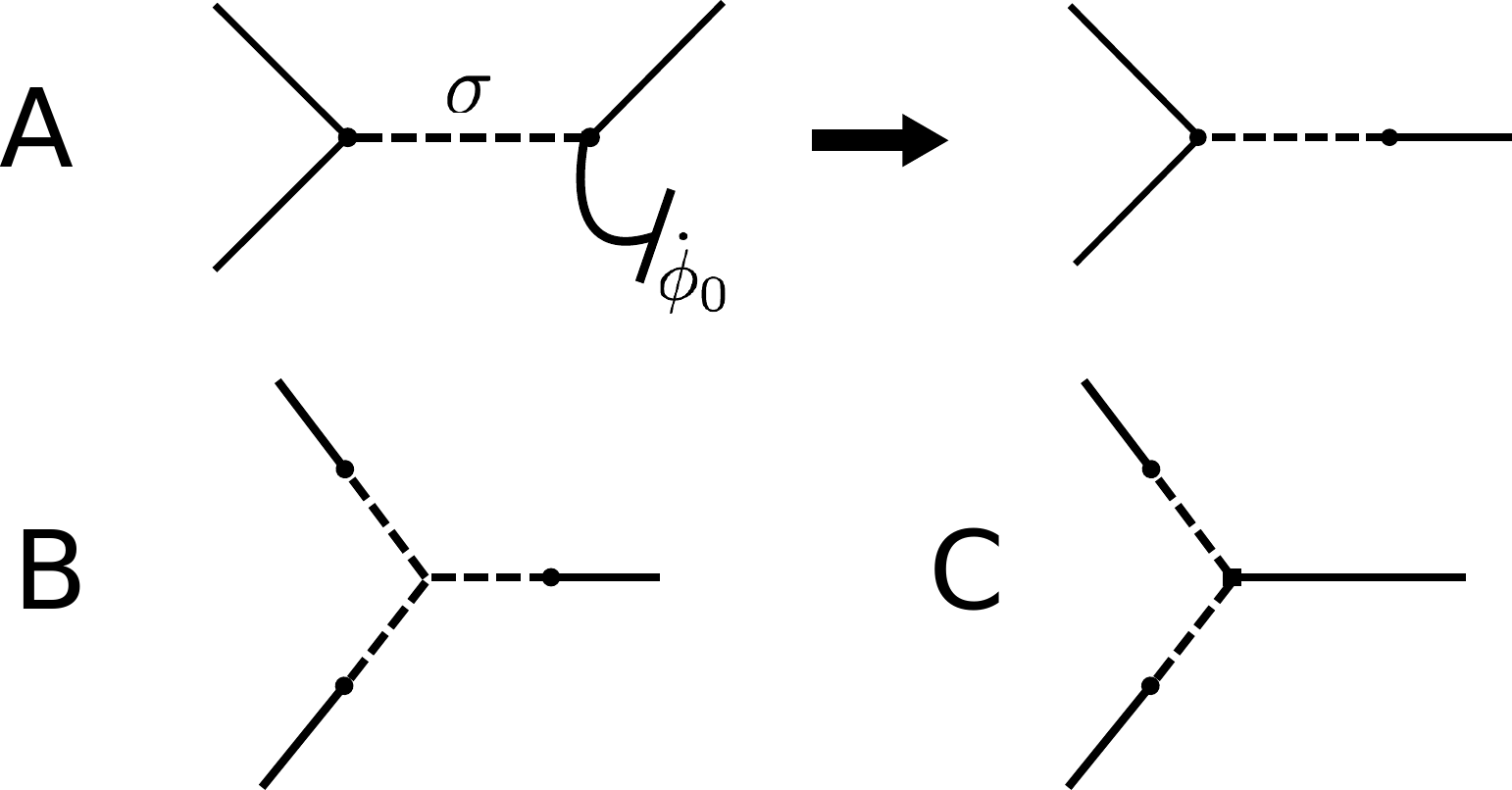}
}
\caption{Simplest tree-level bispectrum Feynman diagrams for QSFI. Type A diagrams, where one leg is taken to be the background field, were computed in \cite{Arkani-Hamed:2015bza,Chen:2014cwa}. Type B diagrams were computed in \cite{Chen:2014cwa}. Diagrams of type C were included in the general EFT studies in \cite{Noumi:2012vr,Baumann:2011nk}.}
\label{fig:feynmans1}
\end{figure}

The model proposed in Ref.~\cite{Chen:2009we} contains a heavy scalar field $\sigma$ producing non-Gaussianities from a self-interaction $\sigma^3$ vertex. As the massive field is not constrained by the approximate shift symmetry of the inflaton field, it can have strong interactions. The non-Gaussianities are then mediated to the curvature perturbation $\zeta$ by a quadratic coupling of the form $\dot{\zeta}\sigma$. The process is described by diagram B of Fig.~\ref{fig:feynmans1}. The quadratic coupling in this model is not put in by hand, but generated by a curved trajectory of the background field with constant radius of curvature (see Ref.~\cite{Chen:2009we} for details), allowing a naturally efficient coupling of the two fields.

Another scattering process has been studied in detail recently \cite{Arkani-Hamed:2015bza,Chen:2014cwa} and is illustrated in diagram A. A vertex of the form $\dot{\zeta}^2\sigma$ or $(\partial_{\bf x}\zeta)^2\sigma$ creates a massive exchange particle $\sigma$. The particle is then converted back by a quadratic coupling of form $\dot{\zeta}\sigma$. The conservative value for the quadratic coupling in the absence of a non-trivial background evolution, is obtained by setting one of the legs of the cubic vertex to the background field proportional to the slow-roll parameter. This type of interactions can be realized as a consequence of turning trajectory \cite{Chen:2009zp}.

A variety of other scattering processes can occur (see for example diagram C) and have been discussed in \cite{Chen:2009zp,Baumann:2011nk,Noumi:2012vr,Lee:2016vti}. In the following we briefly summarize the key scaling behavior in the squeezed limit, which is a model-independent result and allows us to perform a more general analysis in the following sections.

Information about the mass spectrum is encoded in the power of the momentum ratio in the squeezed-limit bispectrum, irrespective of the details of the scattering processes. A key distinction can already be seen at the level of the mode function of a massive particle. Outside of the horizon, the amplitude of a particle with mass $m$ decays as $(\pm \tau)^{3/2\pm i\mu}$ where $\mu=\sqrt{\frac{m^2}{H^2}-\frac{9}{4}}$, $H$ is the Hubble parameter, and $\tau$ is the conformal time. This immediately suggests three qualitatively different mass ranges. For $m/H \approx 0$, the particle does not decay and the phenomenology will be described by what is usually called multi-field inflation. For $0 < m/H \leq 3/2$, the field decays, with a rate determined by the mass parameter. Finally, for $m/H > 3/2$ the field decays while experiencing oscillations as a function of $\tau$. These behaviors have been shown to lead to a characteristic scaling of the squeezed limit of the bispectrum given by
\be
S_{\rm squeezed} \propto \left( \frac{k_{\rm long}}{k_{\rm short}} \right)^{1/2 \pm i\mu}.
\ee
This form reproduces the power-law of the squeezed limit of the local shape when $m/H \approx 0$ and interpolates between the local and equilateral shapes up to $m/H = 3/2$. For higher masses, it becomes an oscillatory function of $k_{\rm long}/k_{\rm short}$.

In principle, the bispectrum receives contributions from all interaction vertices and quadratic mixing vertices that can be generated from $\zeta$, $\sigma$ and their derivatives.
The physical arguments that lead to the above power-law/oscillatory behavior do not depend on the precise couplings of the theory, making it quite generic. This is the most important point of the cosmological-collider analogy \cite{Arkani-Hamed:2015bza}.
Therefore, the squeezed limit of the bispectrum, and more generally soft limits of non-Gaussianities, can be considered as a particle detector for massive particles.

The possible amplitudes of these bispectra span a wide range of values, and strongly depend on the nature of couplings in the model \cite{Chen:2009zp,Baumann:2011nk,Chen:2015lza,Lee:2016vti}.
It is the purpose of this paper to investigate what sizes of the bispectrum may be measurable and the detectability of heavy particles in the futurist 21-cm experiments. We will also discuss the implications on model building.

Finally, we note that for large masses $m>\CO(H)$ the signal acquires a Boltzmann suppression $e^{-\pi \mu}$ \cite{Arkani-Hamed:2015bza}, since particles that are too massive cannot be produced thermally during inflation. This factor limits the mass range of this primordial ``particle detector". So we focus on particles whose masses are not too much heavier than $H$.

Previous works \cite{Sefusatti:2012ye,Norena:2012yi} have considered the detectability of these signals with the cosmic microwave background and large-scale structure in the mass range $m/H \leq 3/2$ (which corresponds to the power-law behavior in the squeezed limit). In this work, we consider the 21-cm experiments and study both the range $m/H \leq 3/2$ and $m/H > 3/2$ (i.e. both the power-law and oscillatory behaviors).

\subsection{The primordial shape function}

The primordial shape function has been evaluated in \cite{Arkani-Hamed:2015bza,Chen:2015lza} for diagrams of type A in Fig.~\ref{fig:feynmans1}. Except for the squeezed limit shape, there is no full analytic expression available, but partial results including the full shape that contains the clock signal have been been provided \cite{Chen:2015lza}. Ref.~\cite{Chen:2015lza} also showed how to efficiently evaluate the necessary integrals using Wick rotations. We review the derivation of the shape from the in-in formalism and details of the numerical evaluation of the integrals in Appendix~\ref{sec:primordialcalc}.

In the present analysis we work in terms of primordial curvature perturbations $\zeta$, with power spectrum
\be
\label{eq:pprim}
\langle \zeta(\mathbf k_1) \zeta(\mathbf k_2) \rangle \equiv (2 \pi)^3 \delta_D(\mathbf{k}_{12}) P_\zeta(k),
\ee
where $\mathbf k_{12} \equiv \mathbf k_{1} + \mathbf k_{2}$, $P_\zeta(k) =A/k^3$, with $A = 2 \pi^2 A_s$, and $A_s$ has the Planck best-fit value $A_s=2.2 \times 10^{-9}$ \cite{Ade:2015lrj}, and we have ignored the small running of the power spectrum. We define the shape $S(k_1,k_2,k_3)$ of the three-point function as
\bea
\langle \zeta^3 \rangle \equiv (2\pi)^3 \delta_D(\mathbf{k}_{123}) \frac{A^2}{(k_1k_2k_3)^2} S(k_1,k_2,k_3).
\eea
We also define the bispectrum quantity $B_{\rm prim}$ through
\be
\label{eq:bprim}
\langle \zeta^3 \rangle = B_{\rm prim}(k_1,k_2,k_3) (2\pi)^3  \delta_D(\mathbf{k}_{123}) ,
\ee
Namely, $B_{\rm prim}(k_1,k_2,k_3)  = A^2 S(k_1,k_2,k_3) /(k_1^2 k_2^2k_3^2)$. In Appendix \ref{sec:bispectrumNorm} we remind the reader of the normalization used throughout the literature.

\begin{figure*}
\resizebox{0.7\hsize}{!}{
\includegraphics{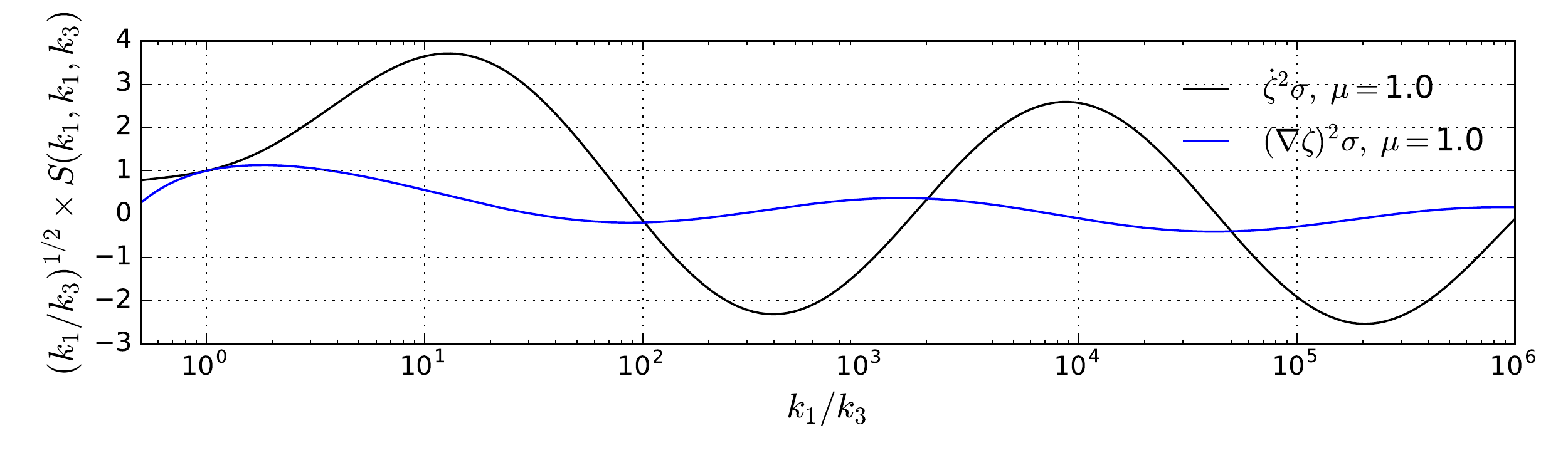}
}
\resizebox{0.7\hsize}{!}{
\includegraphics{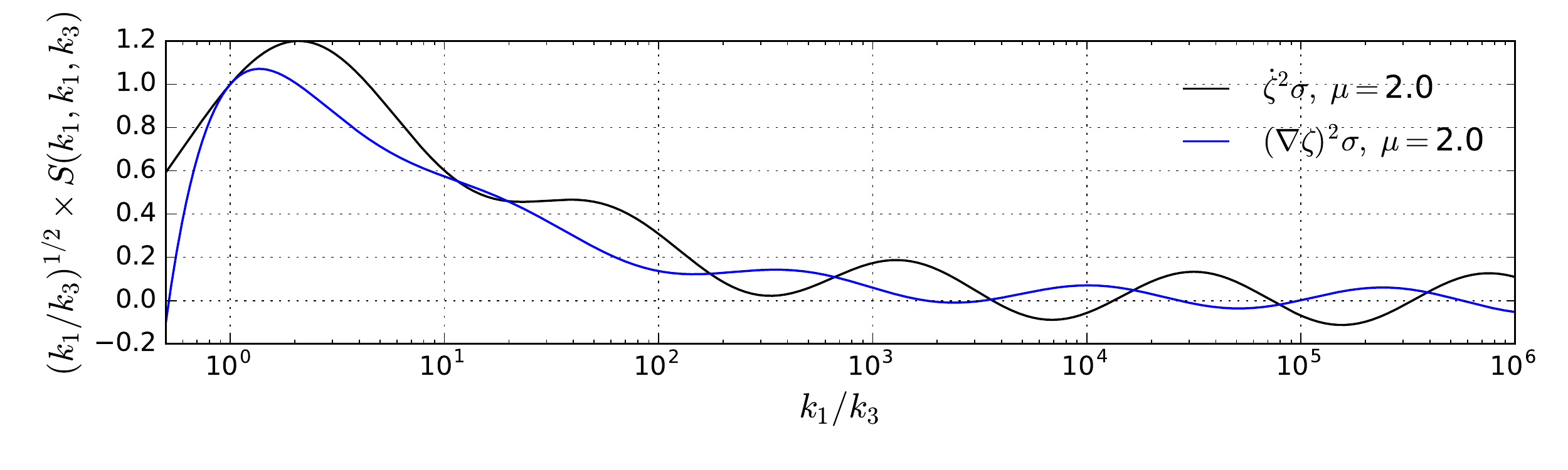}
}
\resizebox{0.7\hsize}{!}{
\includegraphics{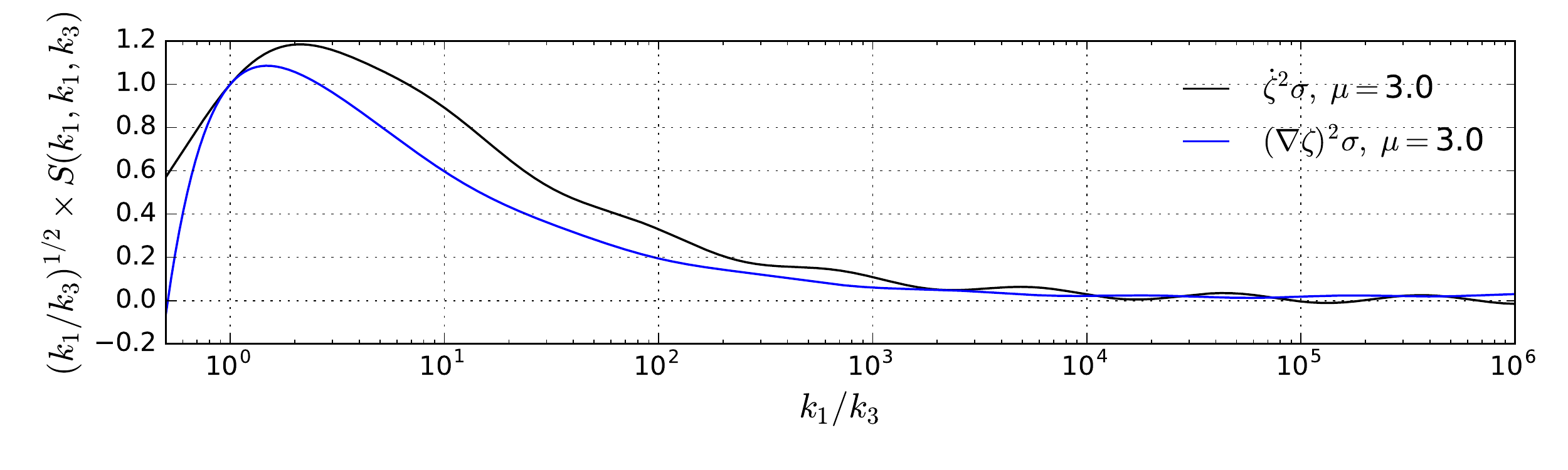}
}
\caption{Bispectrum shape function for different values of $\mu$ for two different couplings, $\dot{\zeta}^2 \sigma$ and $(\nabla \zeta)^2 \sigma$, normalized to unity at $k_1/k_3=1$. The plot contains the full bispectrum, including all terms and permutations. Our results for $\dot{\zeta}^2 \sigma$ are in exact agreement with the results in Ref.~\cite{Chen:2015lza} Fig.~6. Note that the full shape includes both the clock signal (non-analytic in momentum) and non-clock signal (analytic in momentum). The former is Boltzmann-suppressed at large $\mu$, while the latter is power-law-suppressed. In the last two figures, the normalization at the equilateral point is dominated by the latter.}
\label{fig:bispec1}
\end{figure*}

The primordial shape function from particle interactions of the form above can generally be expressed in the form
\bal
\label{eq:chenshape1}
S(k_1,k_2,k_3) =
f_{\rm NL}
&
\left[
\frac{k_1 k_2}{k_3^2} I\left(\frac{k_1+k_2}{k_3}\right) + \frac{k_1 k_3}{k_2^2} I\left(\frac{k_1+k_3}{k_2}\right)
\right.
\nonumber \\
& \left. + \frac{k_2 k_3}{k_1^2} I\left(\frac{k_2+k_3}{k_1}\right)
\right]
~,
\eal
where $I$ is an integral over Hankel functions, which oscillates for squeezed triangles if $m>3H/2$, and $f_{\rm NL}$ is the amplitude of non-Gaussianities. We plot the result for different mass parameters $\mu$ in Fig.~\ref{fig:bispec1}, with an arbitrary scaling to emphasize the oscillations.

\subsection{Degeneracy with other shapes}

\begin{figure}
\resizebox{1\hsize}{!}{
\includegraphics{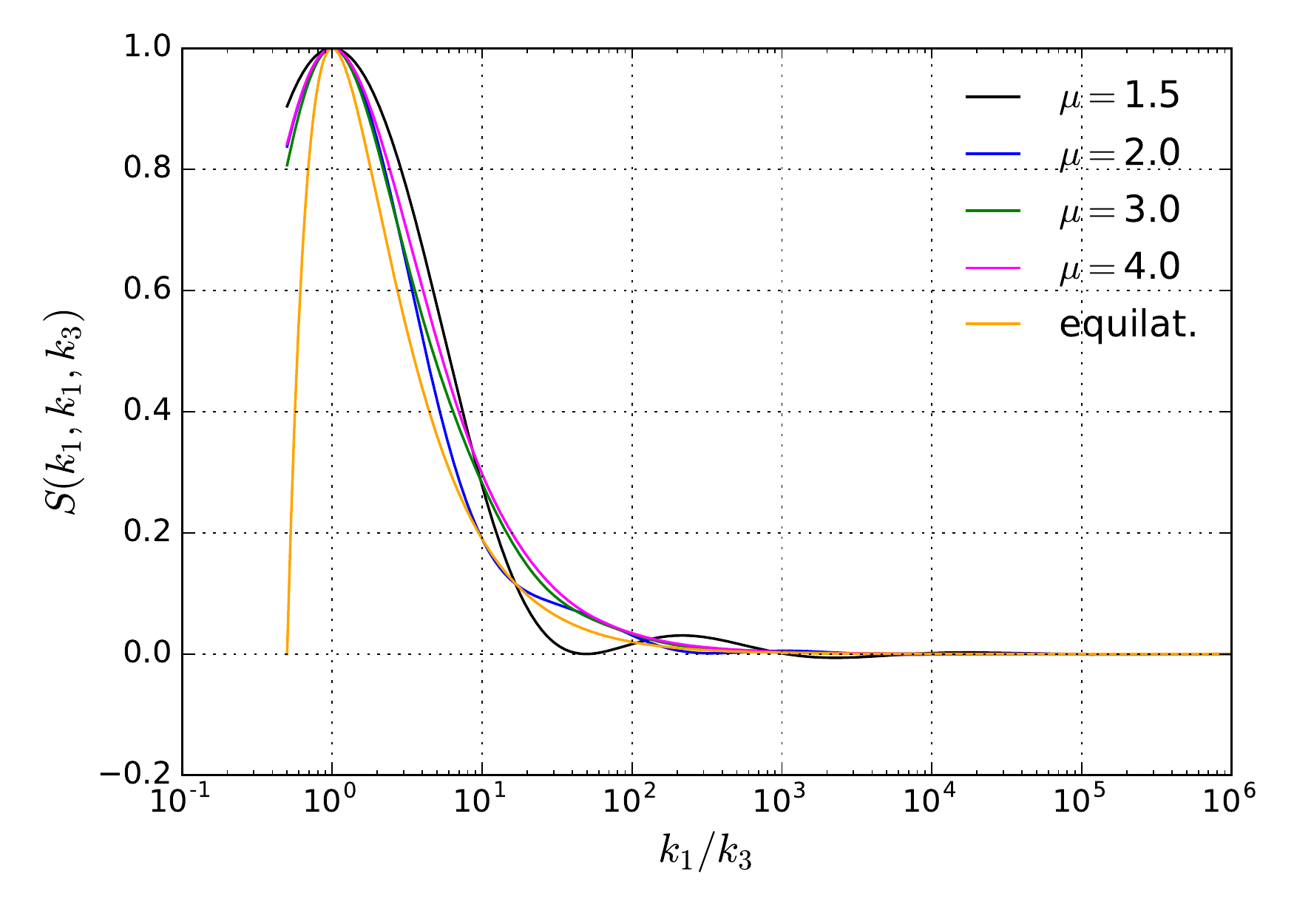}
}
\caption{Primordial shape function on the axis $k_1=k_2$, $k_3=1$ Mpc$^{-1}$, normalized to $1$ for $k_1=1$ Mpc$^{-1}$.}
\label{fig:shapefct}
\end{figure}

Before we calculate the shape overlaps for different values of $\mu$, we examine the shape function on the axis $k_1=k_2$, $k_3=1$ Mpc$^{-1}$ shown in Fig.~\ref{fig:shapefct}. It is clear that this shape peaks for equilateral and flattened configurations. The interesting oscillatory signal is very small for $\mu >1$, so that high $\mu$ are essentially identical. We therefore expect a large overlap with the equilateral shape (yellow curve) and a large cross-correlation for different $\mu$. We will now quantify these statements.

\begin{figure}[t!]
\resizebox{0.8\hsize}{!}{
\includegraphics{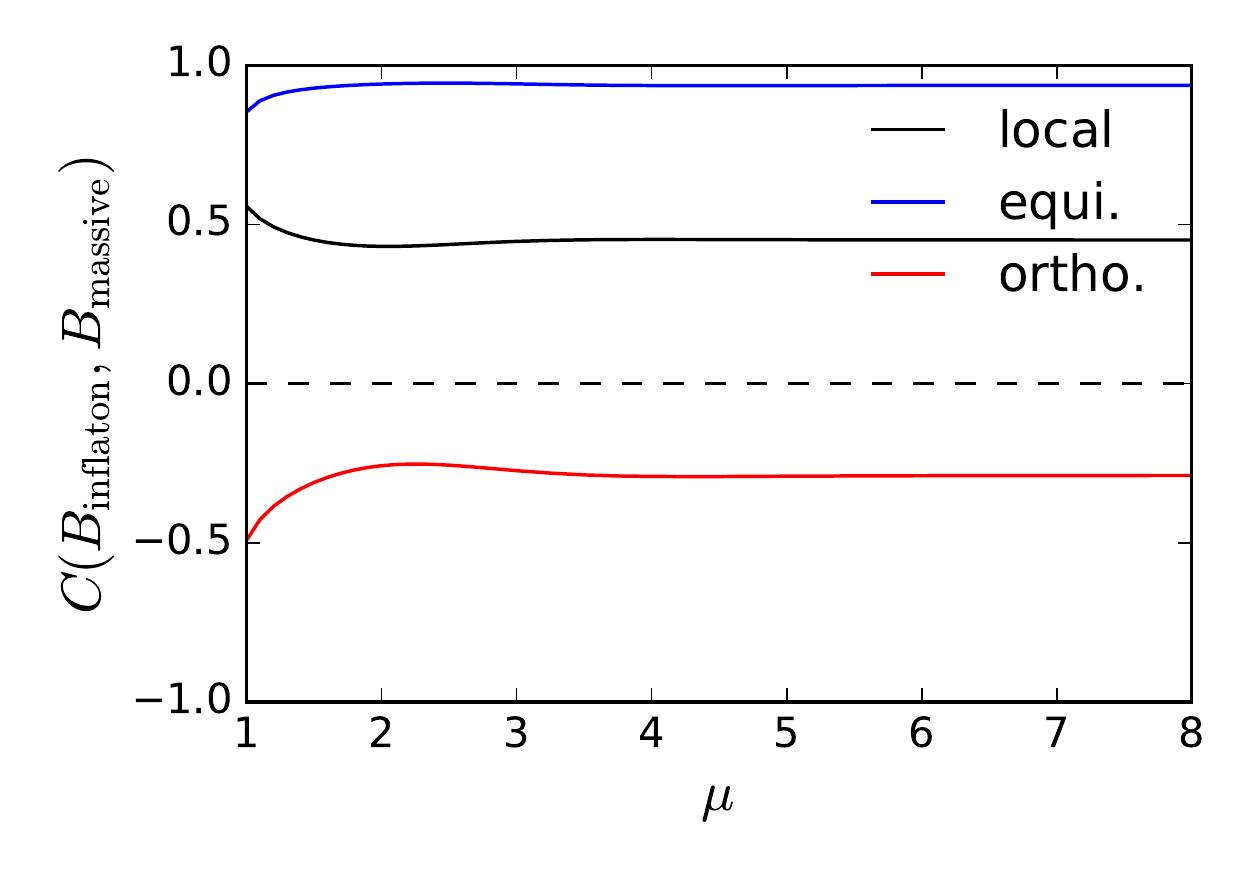}
}
\resizebox{0.7\hsize}{!}{
\includegraphics{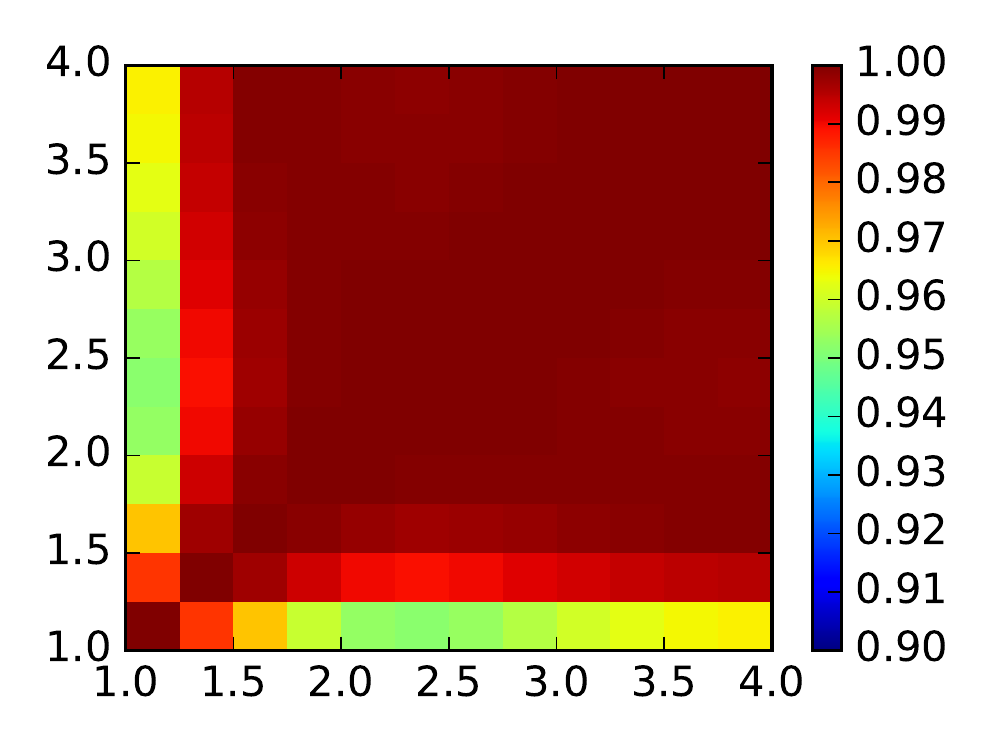}
}
\caption{Top: Primordial bispectrum correlation Eq.~\eqref{eq:innerprod} between the full bispectrum shape of \cite{Chen:2015lza} and the standard inflation shapes as a function of $\mu$. Bottom: Primordial shape auto-correlation of the full bispectrum shape of \cite{Chen:2015lza} as a function of $\mu$. All correlators are very large, showing the small difference between different frequencies. }
\label{fig:corr1}
\end{figure}

The correlation between two different bispectra $B_i(k_1,k_2,k_3)$ and $B_j(k_1,k_2,k_3)$ is given by \cite{Babich:2004gb,Fergusson:2010ia}
\begin{align}\label{eq:innerprod}
\mathcal{C}(B_i,B_j)=\frac{\langle B_i, \, B_j\rangle}{\sqrt{\langle B_i, \, B_i\rangle \langle B_j, \, B_j\rangle}}\,,
\end{align}
where the signal-to-noise-weighted inner product $\left<\right>$ is defined as
\be
\langle B_i, \, B_j\rangle &~\equiv~&
\frac{V}{8 \pi^4}\int_{\mathcal{V}_B}  \prod dk_i \times \nonumber \\
&&\frac{ k_1 k_2 k_3 \,B_i(k_1,k_2,k_3)\,B_j(k_1,k_2,k_3)}{P(k_1)P(k_2)P(k_3)}.\nonumber \\
\label{eq:InnerProd}
\ee
For a scale-invariant spectrum this volume integral simplifies to a 2-dimensional integral. We calculate the auto-correlation of the shape with different mass parameters, shown in Fig.~\ref{fig:corr1} (bottom).  The oscillatory contribution that ``measures'' the mass is strongly suppressed at large $\mu$, leading to a large shape overlap and a poorly-determined mass parameter. For comparison we calculate the overlap with the standard local, equilateral and orthogonal non-Gaussianity shapes in Fig.~\ref{fig:corr1} (top). It is clear that given the large overlap with the equilateral shape, a dedicated search with CMB data for this particular shape will most likely yield a null result given that current bounds on equilateral non-Gaussianity are consistent with zero \cite{PlanckNGs2015} and forecasted constraints will improve by at most a factor of 3 \cite{CMBS4ScienceBook}. This motivates us to propose a template that captures the effect we are aiming to measure and consider more futuristic cosmological observables.

\section{Clock signal templates}\label{sec:template}

In this section we aim to construct bispectrum templates which are maximally sensitive to the signatures sourced by heavy particles. Based on our review and analyses on the cosmological collider signals in the previous section, we would like to achieve two goals in this construction. First, as we pointed-out, although the amplitude of the bispectrum is highly model-dependent, the oscillatory/power-law behavior in the squeezed limit encodes the particle spectrum model-independently; therefore, we would like to construct templates that capture this model-independent behavior while keeping the amplitude as a free parameter. Second, as we have shown, the squeezed-limit of bispectrum is only a small part of the full bispectrum signal, and the full bispectrum has large overlap with the usual equilateral bispectrum which does not contain the particle spectrum information; therefore we would like to remove such ``contamination" from our template.

\subsection{A template for clock signals ($m>3H/2$ case)}

We would like to find a shape function that indicates the presence of heavy particles, without knowledge of the precise inflationary interactions. In the last section we have seen that the shape arising from massive particles has a large overlap with other shapes, as well as with the signal for other masses. To obtain a detection that cannot be confused with self-interactions of the inflaton (which give rise to local or equilateral non-Gaussianities), one has to focus on the oscillatory part. Following Ref.~\cite{Chen:2015lza}, we call the oscillatory contribution the clock signal, and the rest the non-clock signal. The inflationary clock signal has been shown to take the following model-independent form \cite{Arkani-Hamed:2015bza},
\be
S \propto \left( \frac{k_{\rm long}}{k_{\rm short}} \right)^{1/2 \pm i \mu}.
\ee
Motivated by this behavior, as well as the symmetry of Eq.~\eqref{eq:chenshape1}, we therefore propose a template of the form\footnote{With respect to the first preprint version of this paper we have changed the normalization of this shape so that the amplitude in the squeezed limit matches that of the non-oscillating shape to be discussed below in Eq. \eqref{eq:nushape} for $\nu=0$ (the upper mass limit). This choice makes signal comparisons between the two shapes more intuitive.}
\bal
\label{eq:template1}
S^{\mathrm{clock}}(k_1,k_2,k_3) =&
f_{\rm NL} \frac{3^{7/2}}{2}
A\left(\alpha_{123} \right) \left(\alpha_{123} \right)^{-1/2} \times \nonumber \\
&\sin\left(\mu \ln \left(\frac{\alpha_{123}}{2}\right) +\delta \right) + 2\,\mathrm{perm}, \nonumber \\
\eal
where $\alpha_{123} = \frac{k_1+k_2}{k_3}$ and where $\delta$ is a calculable but model-dependent phase. Here $A(\alpha)$ is a window function meant to remove equilateral contributions. We will consider two different window functions, a smooth generalized Gaussian of the form
\be
A_G(\alpha_{123}) = 1-\exp^{ -\left(\frac{\alpha_{123}-1}{a}\right)^{b}},
\ee
and a sharp cutoff
\be
A_H(\alpha_{123}) = \mathcal{H} (\alpha_{123} - \alpha_0),
\ee
where $\mathcal{H}$ is the Heaviside step function. We discuss the choice of the window function and its parameters $a,b$ or $\alpha_0$ next.

\begin{figure*}
\resizebox{0.7\hsize}{!}{
\includegraphics{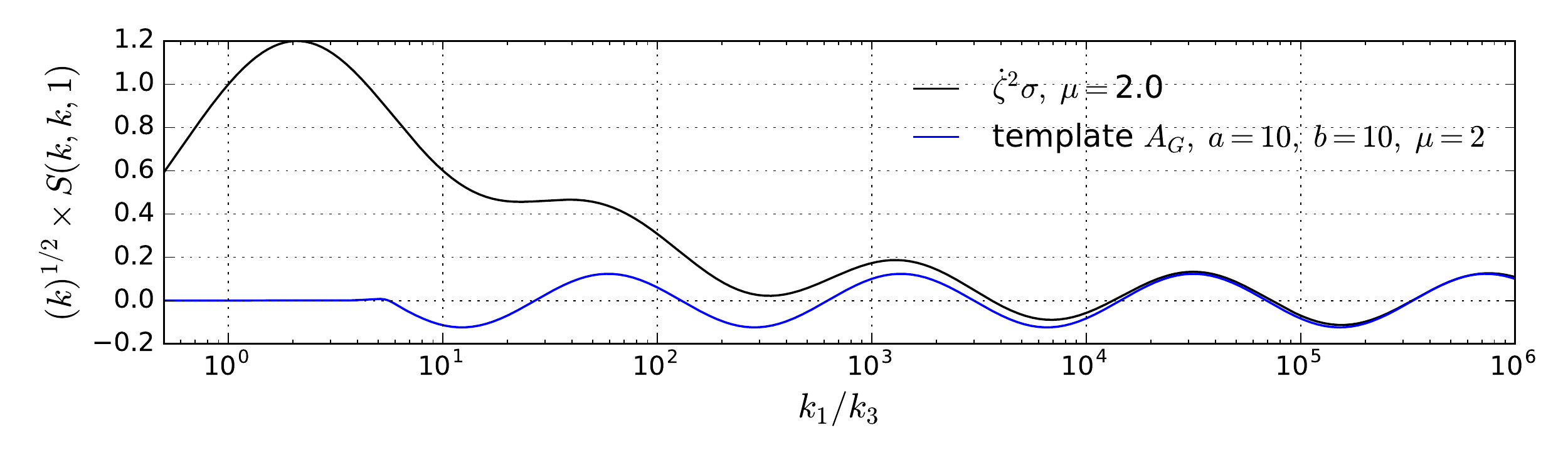}
}
\caption{Template Eq.~\eqref{eq:template1} compared to the full numerical result for $\mu=2$. All permutations are included.}
\label{fig:template1}
\end{figure*}

While the proposed template does not capture the shape perfectly, it should give a good tool for forecasting the crucial clock signal, allowing us to detect massive particles during inflation. The non-separability of the template need not concern us here, as we are not performing a CMB analysis, although factorizability of an oscillatory bispectrum can be achieved as in \cite{OptimalEstimator2014,nonBDbispectrum2015}. The form of the template is shown in Fig.~\ref{fig:template1} where we adjusted the amplitude so that it asymptotes to the numerical result. Given enough oscillations in the observed volume these templates are approximately orthogonal, and one can search for multi particle contributions of the form
\be
S = \sum_i a_i \;S^{\mathrm{clock}}_{\mu_i}(k_1,k_2,k_3),
\ee
where we assumed that the interaction between these particles can be neglected. We emphasize that it is the oscillatory clock signal which is approximately orthogonal, not the sum of the clock and non-clock (equilateral type) signal.

\subsection{Window function and correlators of the template}

The signal-to-noise ratio of the template given an underlying amplitude $f_{\rm NL}$, as well as the overlap of the template with other shapes, depends sensitively on the chosen window function to suppress the equilateral contribution. This choice is influenced by two competing effects. On the one hand, we would like to keep as much of the phase space $(k_1,k_2,k_3)$ as possible, to get the largest amount of signal. On the other hand, the signal is cleanest in the squeezed limit, while in the equilateral region $k_1 \sim k_2 \sim k_3$ there is a large overlap, which prevents us from unambiguously detecting signatures of heavy particles.

\begin{figure*}
\resizebox{0.7\hsize}{!}{
\includegraphics{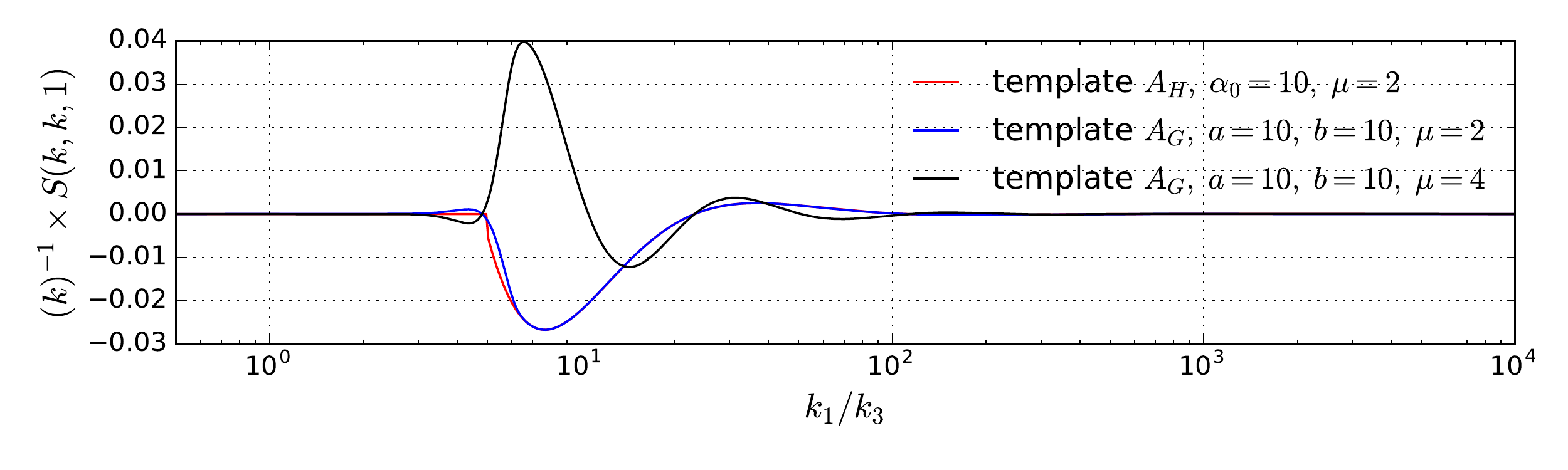}
}
\caption{Characteristics of the generalized Gaussian and the step function envelope. The plot is scaled so that the y-axis indicates the signal-to-noise contribution of triangles with this level of squeezing. The Gaussian $A_G$ has $a=10$ and $b=10$ and the step function $A_H$ has $\alpha_0=10$. For the step function we show two different $\mu$ values. An important lesson from this plot is the importance of the first peak for the signal-to-noise.}
\label{fig:templateenv}
\end{figure*}

To gain intuition into these effects, we show a plot of the template shape as a function of its squeezedness $(k_1/k_3)$, however this time rescaled by powers of the wavenumbers so that the plot is indicative to the signal-to-noise contribution from different values of $(k_1/k_3)$. The plot in Fig.~\ref{fig:templateenv} thus takes into account the number of independent momentum triangles for each configuration, which is inversely proportional to its ``squeezedness". Most of the signal comes from the first oscillation, and highly-squeezed triangles do not noticeably contribute as their number is suppressed by negative powers of $(k_1/k_3)$. Therefore, it is crucial to include as much of the first oscillation as possible to obtain a good estimator. To achieve that goal we have chosen the parameters $a=10$ and $\alpha_0=10$, so that triangles for which $\alpha>10$ contribute to the template, while smaller values are suppressed.
The generalized-Gaussian window function results in a smoother shape and smoother correlators.
We have set its sharpness parameter to be $b=10$, to make sure that equilateral triangles are strongly suppressed.

\begin{figure*}[t!]
\resizebox{1.0\hsize}{!}{
\includegraphics{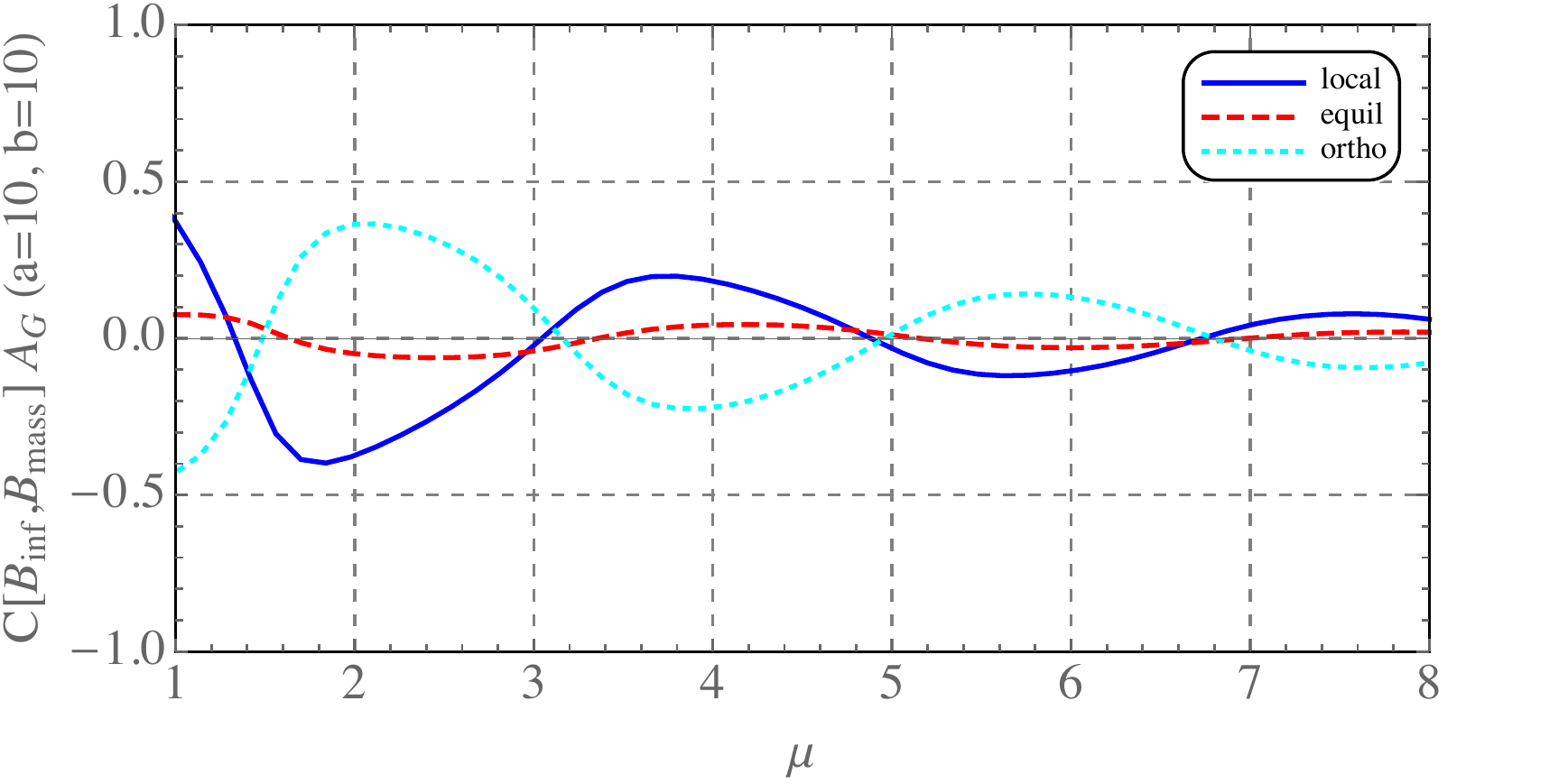}\includegraphics{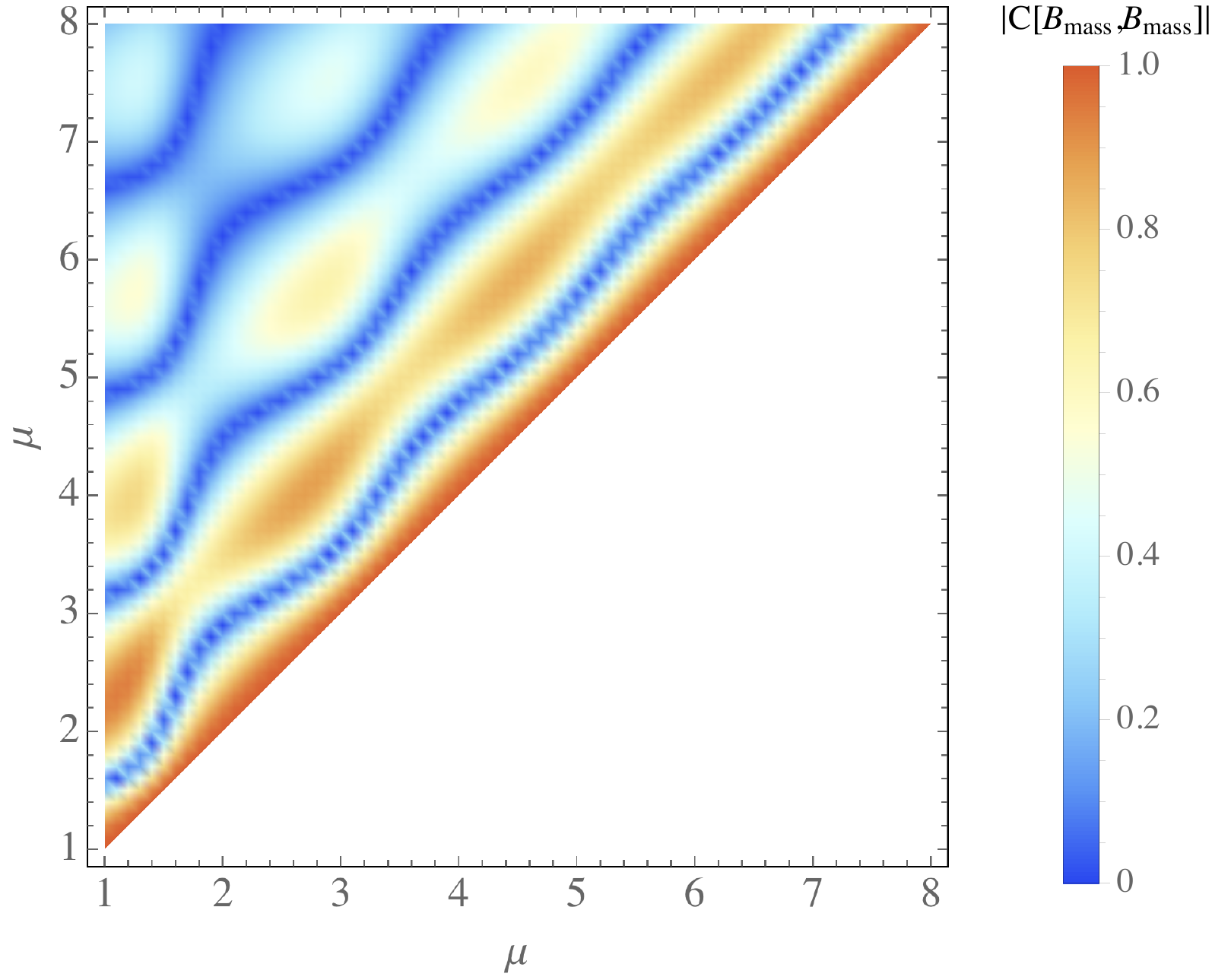}
}
\caption{Left: Primordial bispectrum correlation Eq.~\eqref{eq:innerprod} between the clock template with envelope $A_G$ ($a=b=10$) and the standard inflation shapes as a function of $\mu$. Right: Auto correlation for the clock template between different $\mu$. The auto correlation indicates how precisely different $\mu$ can be discriminated.}
\label{fig:corr2}
\end{figure*}

We plot the primordial correlator of our template (using the generalized-Gaussian envelope with $a=b=10$) with the three standard shapes in Fig.~\ref{fig:corr2} (left). In general, the cutoff induces a phase dependence, which is inherent in the selection of triangles. The equilateral contribution has almost completely been removed. The overlap with the local shape is larger as is appropriate for a shape peaking in the squeezed limit. While at low $\mu$ the overlap with standard shapes is still moderately large, it should be noted that even the local and equilateral shape have a correlator of about $0.4$ in the CMB, so an independent analysis with this correlator is still warranted. In Fig.~\ref{fig:corr2} (right) we also plot the auto correlator of the template shape between different $\mu$, which indicates the precision with which different frequencies can be distinguished effectively.

As we have seen, the overlap depends on the envelope parameter, the phase, and the frequency. For an optimal analysis, the envelope parameter could therefore be chosen frequency and phase dependent. While we are not doing so in our forecast for simplicity, we briefly discuss how the optimal cutoff would be chosen. Let us assume that the underlying bispectrum shape can be decomposed as
\be
S(k_1,k_2,k_3) = a \, S^{\mathrm{equi.}} + b \, S_{p}^{\mathrm{clock}}(k_1,k_2,k_3)
\ee
where the parameter $p$ adjusts the envelope of the clock template, and where the two shapes are normalized so that their integrated signal for $a=b=1$ in the experiment is the same (which requires a change of normalization). As we have seen above, for many models, one expects that equilateral non-Gaussianity would have to be detected first, with large significance, before one could hope to find the clock signal. Next, we assume that we already have detected the equilateral amplitude $a$ with significance $a_\sigma$ sigmas by a future experiment. In this case, one would ask if there is also evidence for the clock signal. The template would then have to be decorrelated so that
\be
\mathcal{C}(S^{\mathrm{equi.}}, S_{p}^{\mathrm{clock}}) \leq \frac{1}{a_\sigma}
\ee
which would determine the choice of $p$.

\subsection{A template for intermediate bispectra ($m<3H/2$ case)}

Because quantum fluctuations of massive fields are most significant if these fields have masses of order $H$, we are most interested in the mass range $|\mu|\lesssim {\cal O}(1)$. This includes both the real and imaginary $\mu$ case. In this section, we present the template for the $m<3H/2$ (i.e.~imaginary $\mu$) case.

For imaginary $\mu$, the template is constructed in \cite{Chen:2009zp}. The oscillatory behavior of the clock signal is analytically continued to a simple power-law behavior, and the shape of the bispectrum lies in between that of the local and equilateral shape, which is called the intermediate shape\footnote{With respect to the first preprint version of this paper we have changed the normalization of this shape to enforce $S(1,1,1)=6$ to faciliate comparison with other shapes below.}:
\bal
\label{eq:nushape}
\hat S^{\rm int} (k_1,k_2,k_3) &=
f_{\rm NL}
\ 2 \ 3^{\frac{7}{2}-3\nu}
\frac{k_1^2+k_2^2+k_3^2}{(k_1+k_2+k_3)^{\frac{7}{2}-3\nu}} (k_1k_2k_3)^{\frac{1}{2}-\nu} \nonumber \\
\eal
where $\nu\equiv\sqrt{(9/4)-(m/H)^2}=-i\mu$ and $0<\nu<3/2$.
In the squeezed limit $k_3\ll k_1=k_2$,
$\hat{S}^{\rm int} \sim (k_3/k_1)^{\frac{1}{2}-\nu}$, in contrast to the local shape $S^{\rm loc.} \sim (k_3/k_1)^{-1}$ and the equilateral shape $S^{\rm equi.} \sim k_3/k_1$. The difference between these shapes is most apparent in the squeezed configuration.
However, the intermediate shape template has large overlap with either the equilateral or local template, especially for $0<\nu<1/2$ due to the equilateral region. For CMB and LSS this is the main obstacle to distinguish $\nu\sim 0$ for $f_{\rm NL} \leq 100$ \cite{Sefusatti:2012ye}.
For 21-cm, we expect to have much more squeezed configurations, and we may be able to afford to have a window function that cuts off the equilateral regions of the template, as we did for the clock signal, e.g. ,
\be
S^{\rm int} (k_1,k_2,k_3) &=& f_{\rm NL} \left( A_G(\alpha_{123})+A_G(\alpha_{231})+A_G(\alpha_{312})
\right) \nonumber \\
&& \hat S^{\rm int}  (k_1,k_2,k_3)
~.
\ee

\subsection{Adding Spin}

A further generalisation can be made to particles with spin, which was shown in Ref.~\cite{Arkani-Hamed:2015bza} to give rise to a squeezed limit
\be
S \propto \left( \frac{k_{\rm long}}{k_{\rm short}} \right)^{1/2 \pm \mu} P_s(\cos \Theta)
\ee
where $\Theta$ is the angle between $k_{\rm long}$ and $k_{\rm short}$.

When adding spin to the massive interacting fields, the bispectra pick up a Legendre term $P_s(\cos \Theta)$ \cite{Arkani-Hamed:2015bza}, with $\Theta$ the angle between the $k_1$ and $k_3$ (+ 2 permutations). To add this effect to our template, we use that
\be
\cos \Theta \equiv \hat{k}_1 \cdot \hat{k}_3 = \frac{k_1^2 +k _3^2-k_2^2}{2k_1k_3}.
\ee
Note that this is already of the factorized form, and Legendre polynomials are just sums of power laws in this quantity, which remain factorized for each spin. We do not consider spin here, and leave this for a future study. See e.g \cite{2016arXiv160705232C} for a recent attempt to constrain the effects of anisotropic non-Gaussianity on galaxy shapes.

\section{21-cm forecast for the template}
\label{sec:21-cm}

Due to spin-orbit coupling, neutral Hydrogen has two differentiated ground states, a singlet and a triplet, where the latter can decay to the former by emitting a photon with a 21-cm wavelength  (or 1400 MHz frequency).
Most baryonic matter in our Universe is in the form of Hydrogen, thus being able to emit and absorb in this line if neutral. Due to the large comoving volume, as well as the high resolution that can be achieved by observing this redshifted 21-cm line, there has been numerous studies into using it for cosmological measurements \cite{Loeb_2004,Bharadwaj_2004,Fuetal2006,Fur2009proc,Pritchard:2011xb}.
Of particular interest is the redshift range $z = 30-100$, also known as the ``dark ages", where the 21-cm line can be observed in absorption against the CMB.

Let us name $n_0$ and $n_1$ the number densities of Hydrogen in the singlet and triplet state, respectively. We can then define the spin temperature as \cite{Wouthuysen_1952,Field_1958}
\be
T_s = T_* / \log\left(\dfrac{3n_0}{n_1}\right),
\ee
where $T_*=0.068\,K$ is the energy corresponding to the 21-cm transition. During the dark ages the spin temperature can be written as
\be
T_s = T_\gamma + \dfrac{C_{10}}{C_{10} + A_{10} \frac{T_{\rm gas}}{T_*}} (T_{\rm gas} - T_\gamma),
\ee
where $A_{10}$ is the Einstein spontaneous-emission coefficient, $C_{10}$ the collisional coefficient, and $T_{\rm gas}$ and $T_\gamma$ are the gas and CMB temperatures, respectively. As long as the Hydrogen spin temperature is lower than that of the CMB there will be absorption of 21-cm photons from the Rayleigh-Jeans tail of the CMB. The optical depth for this absorption is
\be
\tau = \dfrac{3}{32\pi}\dfrac{T_*}{T_{s}} n_{\rm HI} \lambda_{\rm 21}^3 \dfrac{A_{10}}{H(z)+(1+z)\partial_r v_r},
\ee
where $n_{\rm HI}$ is the neutral-Hydrogen number density, $\lambda_{21}$ is the wavelength of the 21-cm transition, $H(z)$ the Hubble parameter, and $\partial_r v_r$  the radial velocity gradient. This will create a temperature contrast with respect to the CMB given by
\be
T_{21} = \tau \dfrac{T_s-T_\gamma}{1+z}.
\ee

The distribution of neutral Hydrogen at high redshift is not perfectly homogeneous. This induces perturbations to the brightness temperature of different patches of sky, and in terms of Fourier modes the temperature fluctuation can be written to linear order as
\be
\delta T^{\rm lin}(\bsk) = (\alpha + \overline{T}_{21} \mu^2) \delta_b(\bsk),
\label{eq:dT21}
\ee
where we have defined $\mu=\bsk \cdot \hat n/k$, and the redshift-dependent functions $\alpha(z)=\mathrm dT_{21}/\mathrm d\delta_b$ and $\overline{T}_{21}(z)$, also known as the global signal, can be found on Ref.~\cite{21cmFNL}. This yields a temperature power spectrum
\be
P_{\delta T}(\bsk) = \left(\alpha + \overline{T}_{21}  \mu^2\right)^2 P_{\delta_b}(k).
\ee

\subsection{The 21-cm bispectrum}

We are interested in the bispectrum of 21-cm fluctuations \cite{Cooray:2006km,Pillepich_2007}. The temperature three-point function can be written
as
\be
\langle \delta T_{21}(\bsk_1) \delta T_{21}(\bsk_2) \delta T_{21}(\bsk_3) \rangle &=& (2 \pi)^3 \delta_{\rm D}(\bsk_1 + \bsk_2 + \bsk_3) \nonumber \\
&& \times B_{\delta T}(\bsk_1, \bsk_2, \bsk_3).
\ee
Given a primordial bispectrum $B_{\rm prim}$, defined in eq.~\eqref{eq:bprim}, one can find the resulting temperature bispectrum as \cite{21cmFNL}
\be
B_{\delta T,\rm prim}(\bsk_1, \bsk_2, \bsk_3) &=&B_{\rm prim}(k_1,k_2,k_3) \nonumber \\
&& \prod_{i=1}^3 [\alpha + \overline{T}_{21} \mu_i^2] \mathcal T(k_i), \nonumber \\
\ee
where $\mu_i \equiv \bsk_i\cdot\hat n/k_i$, and $\mathcal T(k)$ is the linear transfer function for baryons. In the next section we forecast the sensitivity for primordial bispectra parametrized by the shape functions
\be
\label{eq:template3}
S^{\mathrm{clock}}(k_1,k_2,k_3) &=& f_{\rm NL} \frac{3^{7/2}}{2}  A_G\left(\alpha_{123} \right) \left(\alpha_{123} \right)^{-1/2} \nonumber \\
&& \sin\left(\mu \ln \left(\frac{\alpha_{123}}{2}\right) +\delta \right) + 2\,\mathrm{perm}.\nonumber \\
\ee
with envelope $A_G(\alpha_{123}) = 1-\exp^{ -\left(\frac{\alpha_{123}-1}{10}\right)^{10}}$ and $\alpha_{123} = \frac{k_1+k_2}{k_3}$. We will also consider the case with imaginary $\mu = i \nu$, where our template
will be
\be
\label{eq:template4}
S^{\rm int} (k_1,k_2,k_3) &=& f_{\rm NL}
\left[ A_G(\alpha_{123})+A_G(\alpha_{231})+A_G(\alpha_{312})
\right] \nonumber \\
&& \ 2 \ 3^{\frac{7}{2}-3\nu}
\frac{k_1^2+k_2^2+k_3^2}{(k_1+k_2+k_3)^{\frac{7}{2}-3\nu}}
(k_1k_2k_3)^{\frac{1}{2}-\nu} ~.\nonumber \\
\ee

\begin{figure*}[t]
\resizebox{0.8\hsize}{!}{
\includegraphics{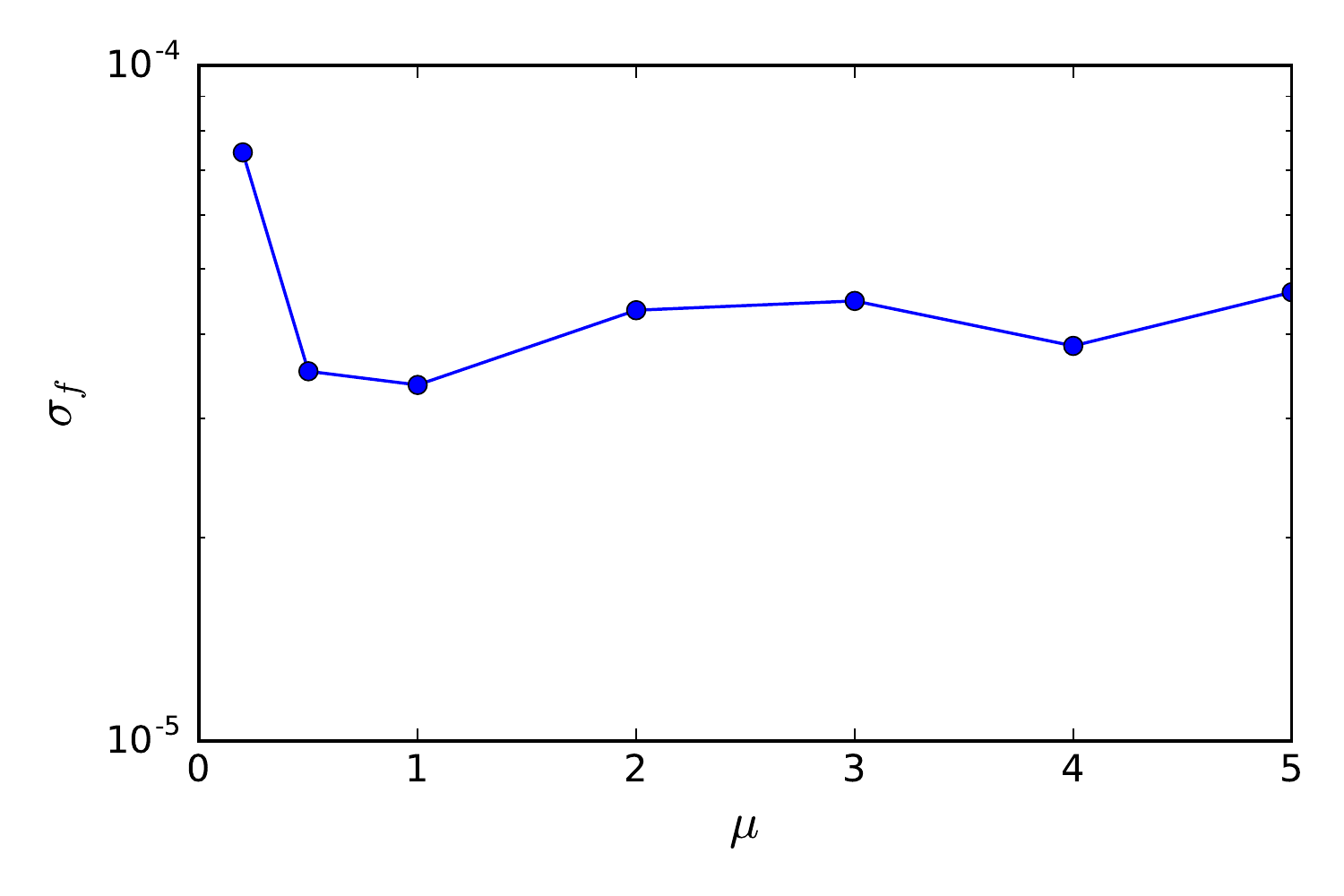}\includegraphics{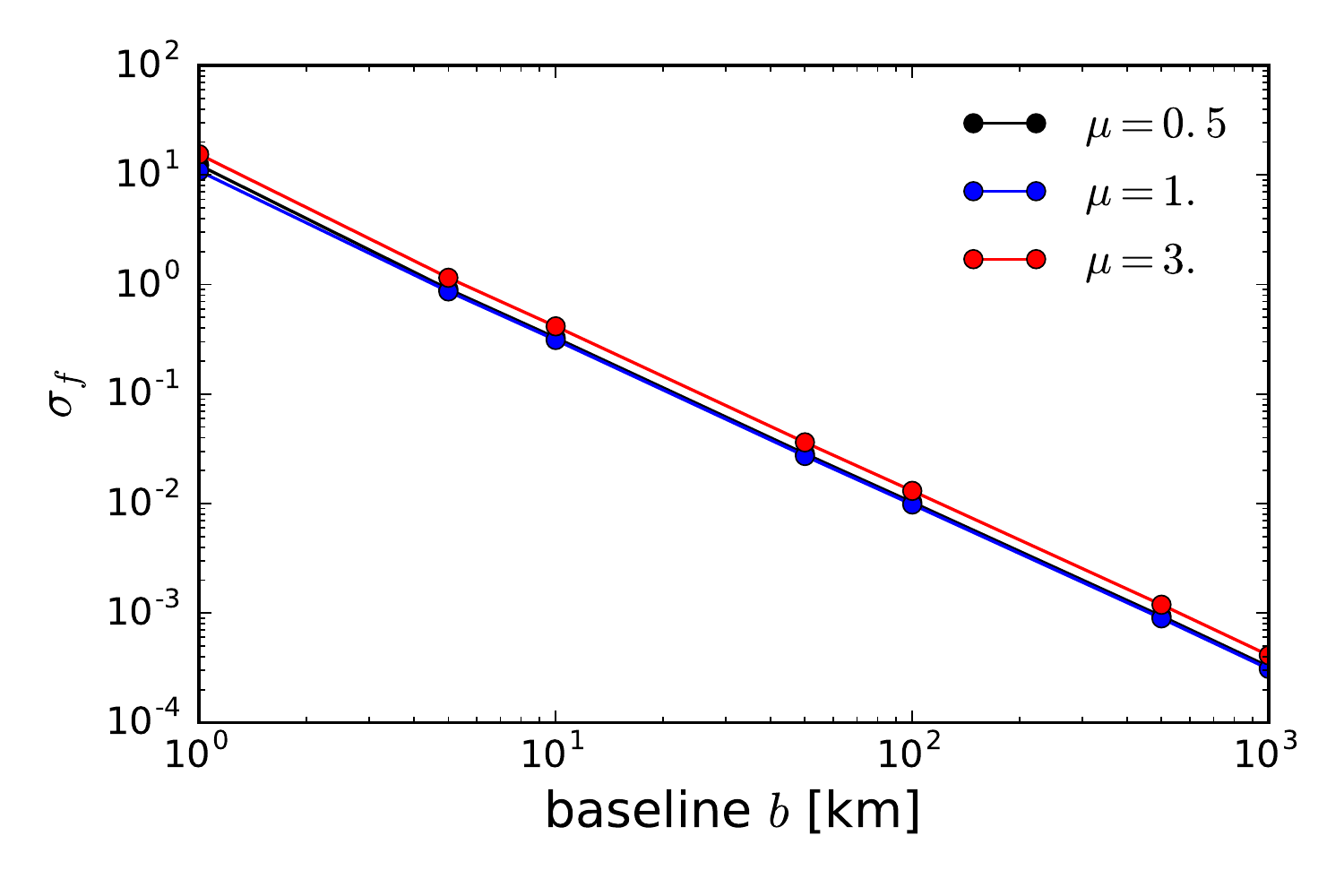}
}
\resizebox{0.8\hsize}{!}{
\includegraphics{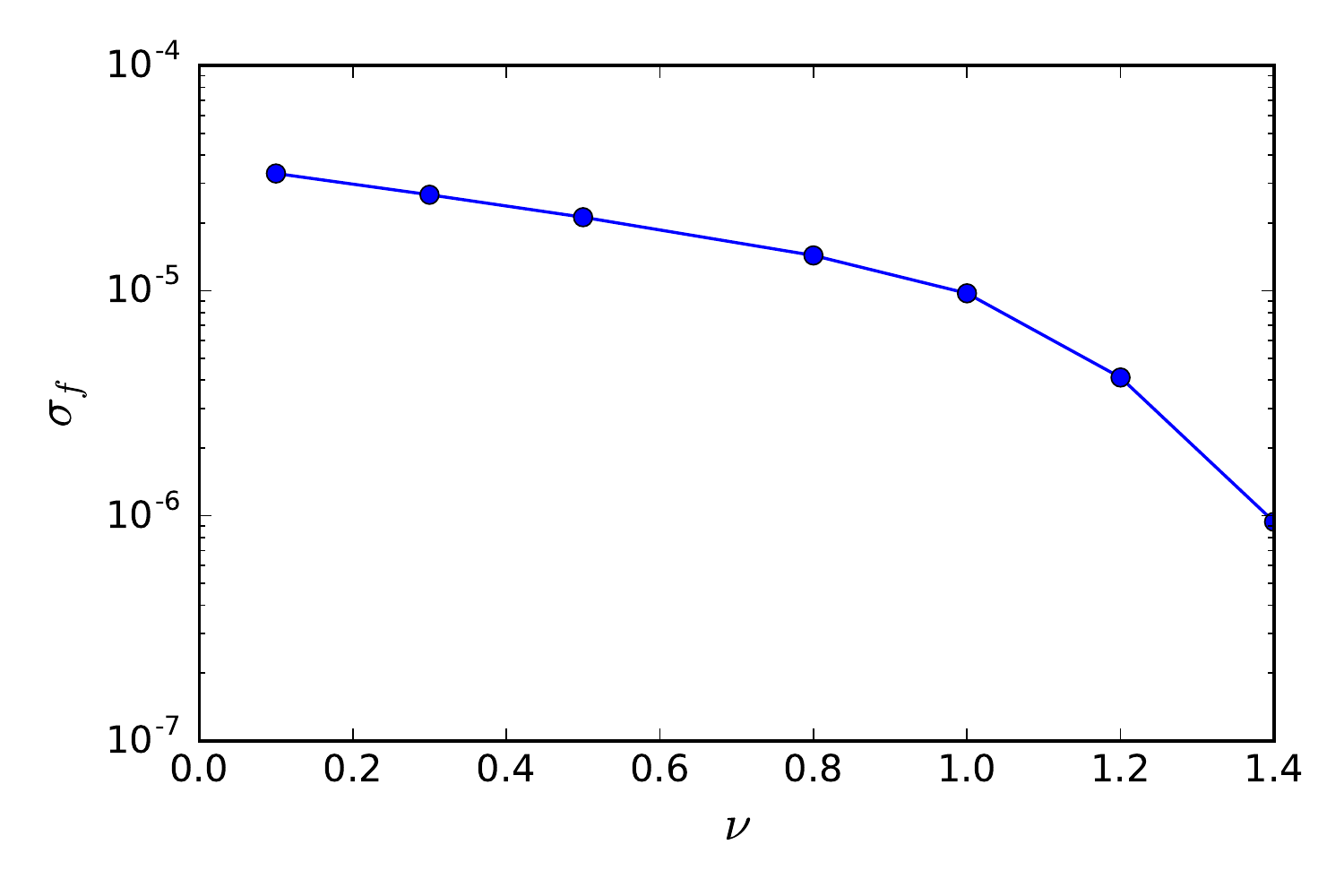}\includegraphics{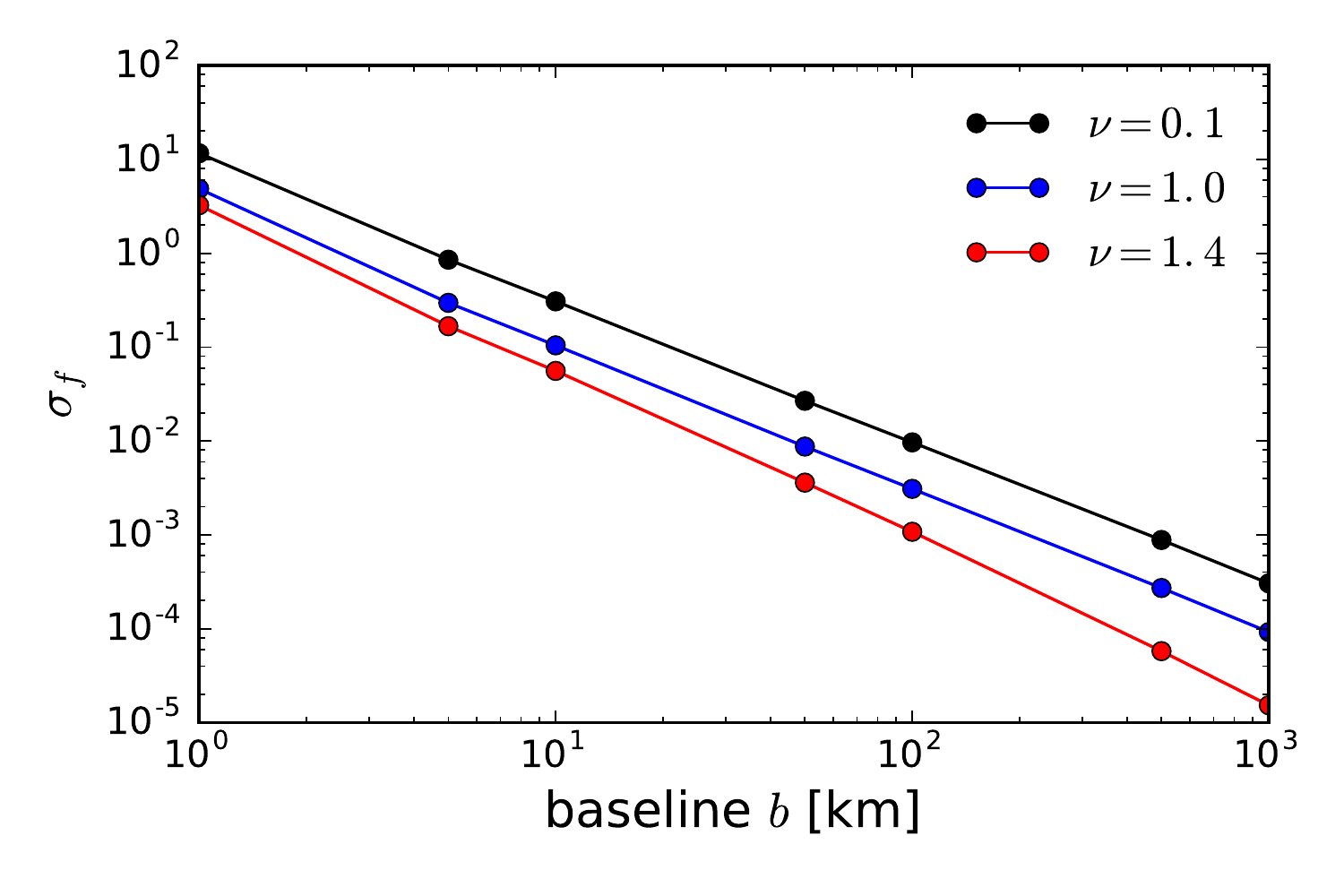}
}
\caption{Top left: Ultimate cosmic variance limited sensitivity for template Eq.~\eqref{eq:template3} with $k_{\rm max}=300$ Mpc$^{-1}$ (Jeans limit) and $30<z<100$ as a function of $\mu$. Top right: Sensitivity as a function of the baseline of a noiseless experiment, assuming the radial resolution matches the angular resolution. Bottom plots: same quantities for template Eq. \eqref{eq:template4} as a function of $\nu$. In all plots the frequency, phase and cosmology is fixed (we study the frequency determination below). Note that even a baseline of 1000km still does not reach the $k_{\rm max}=300$ Mpc$^{-1}$ limit of the left side plots.}
\label{fig:fishermax}
\end{figure*}

\subsection{Secondary non-Gaussianities}
\label{Sec:Secondary}

Due to the slow collapse
of the Hydrogen clouds under gravity, 21-cm fluctuations from the dark ages are not perfectly linear. Gravity, being a non-linear theory,
generates secondary non-Gaussianities. In adddition, the fluctuations of the 21-cm field are a biased
tracer of the underlying density field, causing additional secondary non-Gaussianities.

It was shown in Ref.~\cite{21cmFNL} that these secondary non-Gaussianities dominate the signal, and would
swamp the primordial signal by several orders of magnitude.
There are 21 geometrically distinct shapes of secondary non-Gaussianities
to second order in both the baryon overdensity $\delta_b$, and the velocity perturbation $\delta_v \equiv -(1+z) \partial_r v_r/H(z)$.
For this reason it is necessary to marginalize over these secondaries, in order to find the much-smaller primordial signal.

We can decompose the observed $T_{21}$ bispectrum, with the best-fit secondary bispectrum subtracted, as
\begin{align}
B_{\delta T} &= B_{\delta T,\rm prim} + \sum_{i=1}^{21} \sigma_{c_i} \dfrac{\partial}{\partial c_i}B_{\delta T, {\rm sec},i},
\end{align}
where the secondary bispectra are calculated in Ref.~\cite{21cmFNL},
and shown in our Appendix~\ref{app:secondaries}, and $c_i$ are their amplitudes.

\begin{figure}[t!]
\resizebox{\hsize}{!}{	
	\includegraphics{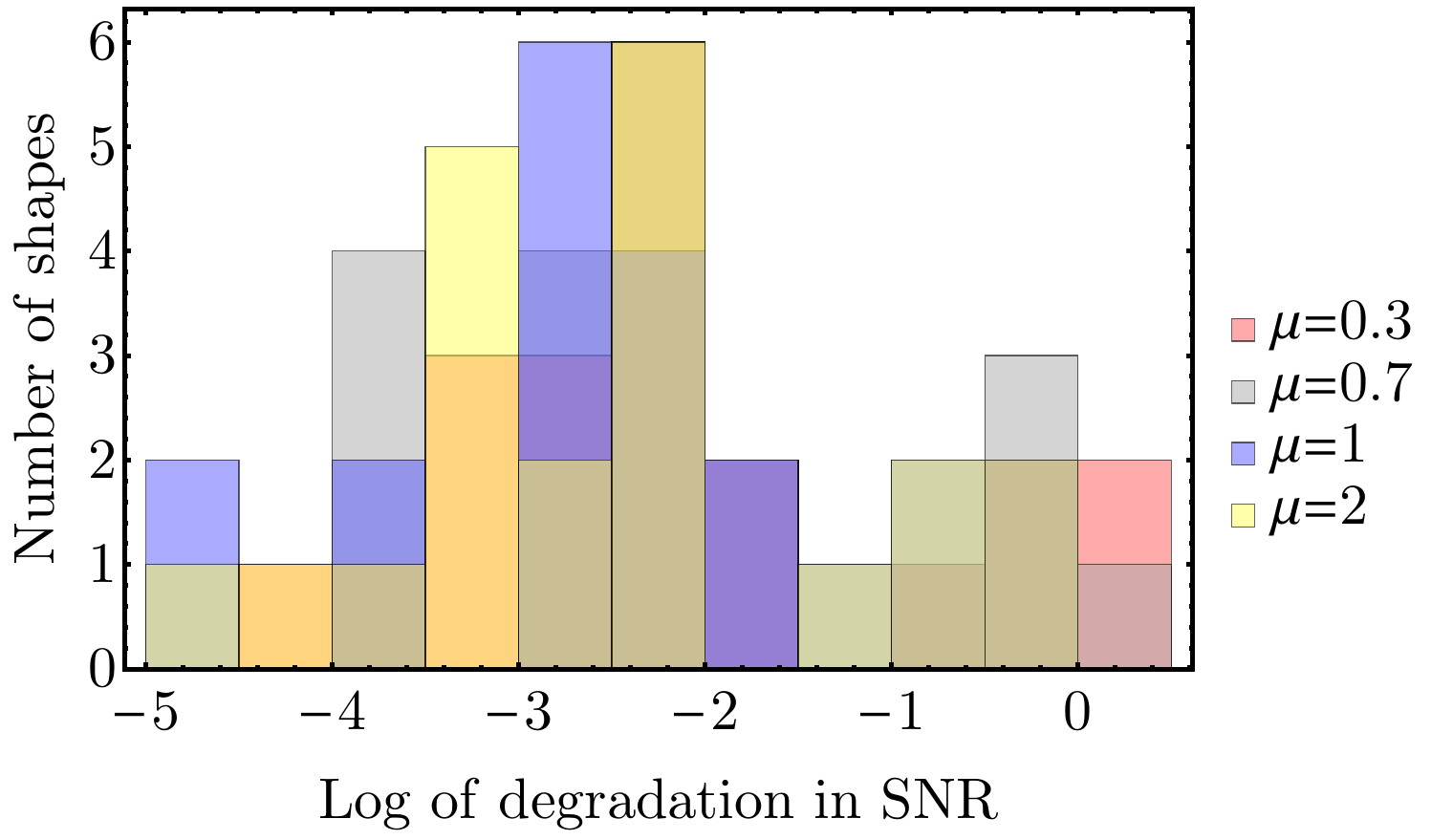}
}
	\caption{Histogram of the (natural logarithm of the) degradation in the SNR, defined as $\sqrt{F^{-1}_{00} F_{00}}-1$, with each of the 21 secondary shapes. In red we show the $\mu=0.3$ case, in gray $\mu=0.7$, in blue $\mu=1$, and in yellow $\mu=2$.}
	\label{fig:degradation1}
\end{figure}

\begin{figure}[t!]
\resizebox{\hsize}{!}{	
	\includegraphics{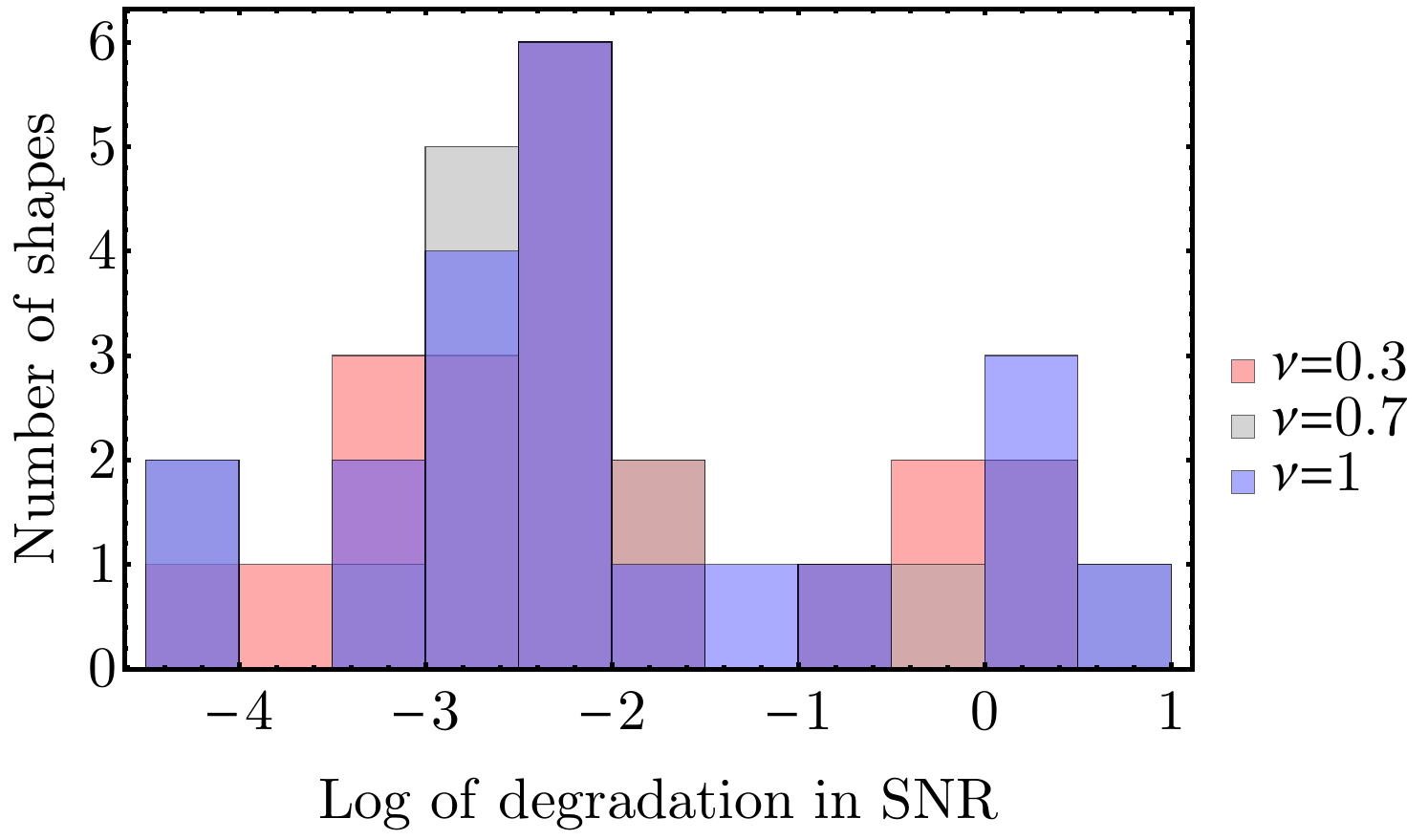}
}
	\caption{Degradation in the SNR, defined as $\sqrt{F^{-1}_{00} F_{00}}-1$, with each of the 21 secondary shapes independently. In red we plot the $\nu=0.3$ case, in gray $\nu=0.7$, in blue $\nu=1$.}
	\label{fig:degradation2}
\end{figure}

We define the signal-to-noise-ratio (SNR) degradation as the amount of signal
lost with respect to the case without secondaries, i.e. $\sqrt{F^{-1}_{00} F_{00}}-1$,
assuming the subindex $0$ corresponds to $f_{\rm NL}$ and the Fisher matrix $F_{ij}$ is computed as in Eq.~\eqref{eq:InnerProd} using $B_{\delta T}(\mathbf k_1,\mathbf k_2,\mathbf k_3)$ and $P_{\delta T}(\mathbf k)$ at $z=50$.
We show a histogram of the degradations in signal to noise when adding each of the 21 independent shapes in Figure~\ref{fig:degradation1} for the $\mu$ case, and in Fig.~\ref{fig:degradation2} for the $\nu$ case.
We find that most secondary shapes are orthogonal to the $\mu$ primordial shapes,
especially for higher $\mu$ values, due to more-rapid oscillations.
In the $\nu$ case, however, there is significant overlap between several secondaries and
the primordial signal, which is to be expected since the $\nu$ shapes interpolate between the local and equilateral templates, already found to be highly correlated with secondaries in Ref.~\cite{21cmFNL}.
We show the total degradation in signal to noise in Tab.~\ref{tab:deg}, where we have marginalized over the 21 secondary shapes simultaneously. All the degradation factors are $\CO (1)$, and are particularly small for the oscillatory ($\mu$) case.
We note that this degradation is driven by a few highly-correlated shapes, so restricting the analysis to 4 linear combinations of the 21 shapes, as done originally in Ref.~\cite{21cmFNL}, would not change results significantly.
We conclude that secondaries would not strongly affect our forecasts, and we ignore them for the rest of this work.

With these approximations, the evaluation of the integral is straightforward. Numerical care must be taken to correctly sample highly-squeezed triangles, which carry a significant part of the signal for this shape.

\begin{table}
	\begin{tabular}{|l|c|}
		\hline
		Model  & SNR Deg.   \\ \hline
		$\mu=0.3$ & 0.99 \\
		$\mu=0.7$ & 0.75 \\
		$\mu=1$ & 0.47 \\
		$\mu=2$ & 0.56 \\
		$\nu=0.3$  & 1.05 \\
		$\nu=0.7$ & 1.92  \\
		$\nu=1$ & 1.44 \\
		\hline
	\end{tabular}
	\caption{Degradation in SNR computed as $\sqrt{F^{-1}_{00} F_{00}}-1$, when considering the 21 secondary shapes of non-Gaussianities simultaneously.}
	\label{tab:deg}
\end{table}

\subsection{Fisher forecast methodology}

A commonly-used, and computationally-simple, method for forecasting the experimental precision for 3D surveys is to work directly in perturbation space, neglecting the spherical geometry of the observation. A recent 21-cm forecast using this method for step features and resonance signals can be found in Ref.~\cite{Xu:2016kwz}. Redshift evolution can be taken into account approximately by binning in several redshift bins with volume $V_i$. The Fisher matrix is (see for example \cite{Scoccimarro:2003wn,Baldauf:2016sjb})
\be
F^b_{ij}= \sum_{z_i} \sum_{T,T'} \frac{\partial B(T,z_i)}{\partial p_i} (C^{-1})_{TT'} \frac{\partial B (T',z_i)}{\partial p_j} \;.
\label{eq:Fisher}
\ee
where $p_i=\{f_{\rm NL},\mu,c_i\}$ are the parameters we want to find, and $T$ is the sum over triangles, i.e.,
\be
\sum_T \equiv \sum_{k_1=k_{\rm min}}^{k_{\rm max}}  \sum_{k_2=k_1}^{k_{\rm max}} \sum_{k_3={\rm max}(k_{\rm{min}},k_2-k_1)}^{k_2} \;,
\ee
and the Gaussian covariance matrix between triangle configurations is
\be
C_{TT'}= \frac{(2\pi)^3}{V_i}   \frac{\pi s_{123}}{dk_1 dk_2 dk_3}  \frac{P(k_1) P(k_2) P(k_3)}{k_1 k_2 k_3}  \delta_{TT'}
\label{eq:CTT}
\ee
with symmetry factor $s_{123}$ (6, 2 or 1 for equilateral, isosceles and general triangles). In this way we can forecast:
(\!{\it i}) the minimum measurable $f_{\rm NL}$ for each $\mu$, when marginalizing over $\mu$, and (\!{\it ii})
the error in $\mu$ given some non-zero value of $f_{\rm NL}$.
From Eq.~\eqref{eq:Fisher} it is clear that $\partial B/\partial \mu \propto f_{\rm NL}$, so $\sigma_\mu \propto 1/f_{\rm NL}$, making the mass of the particles easier to determine with a larger amplitude of non-Gaussianities.

To find the total Fisher matrix we just add the information over all the frequencies $\nu_I$ as
\be
F_{ij}^{\rm tot} = \sum_{\nu_I} F_{ij} (\nu_I) \approx \int_{z_1}^{z_2}  \dfrac{\mathrm dz}{(1+z)^2} \dfrac{\nu_0}{\Delta \nu} F_{ij} (z),
\ee
where $\nu_0=1400$ MHz, and $\Delta\nu$ is the bandwidth. We take as integration limits $z_1=30$ and $z_2=100$,
to represent the observable dark ages. Notice that the bandwidth
factor will cancel with the comoving volume on the denominator of Eq.~\eqref{eq:CTT}.

\subsection{Results}

As an ultimate upper limit of the detectability of our template, we first make a cosmic-variance-limited forecast where $k_{\rm max}=300$ Mpc$^{-1}$ \cite{2014PhRvD..89h3506A}, i.e. around the Jeans limit of the baryon perturbations. The largest observable modes $k_{\rm min}$ are limited by the survey volume only, which we take to cover the redshift range $z=30$ to $z=100$, ignoring possible foreground contamination at low $k$ \cite{Morales:2005qk,Mao:2008ug}.
We show results for this setup in Fig.~\ref{fig:fishermax} (left). As expected, under the simplifying assumption made here, the sensitivity is not strongly depended on the mass parameter $\mu$.
To avoid confusion, we note that the amplitude of our template is a free parameter, while in model building such an amplitude would become increasingly difficult to realize as $\mu$ becomes larger than $\CO(H)$ due to Boltzmann suppression.
In the present analysis we have set the phase to zero. For small values of $\mu$ the amplitude sensitivity is somewhat phase dependent, while for larger $\mu$ the values stabilize. For example, for $\mu=1$, the sensitivity $\sigma_f$ changes with the phase by at most a factor of two. In general, the phase is well determined by the data when the frequency is well determined and vice versa. The phase is computable in a given model and depends on the coupling. Leaving the phase as a free parameter in the data analysis would decrease the frequency resolution, but for the precision of this forecast this can be ignored.

\begin{figure}[t!]
\resizebox{\hsize}{!}{	
\includegraphics{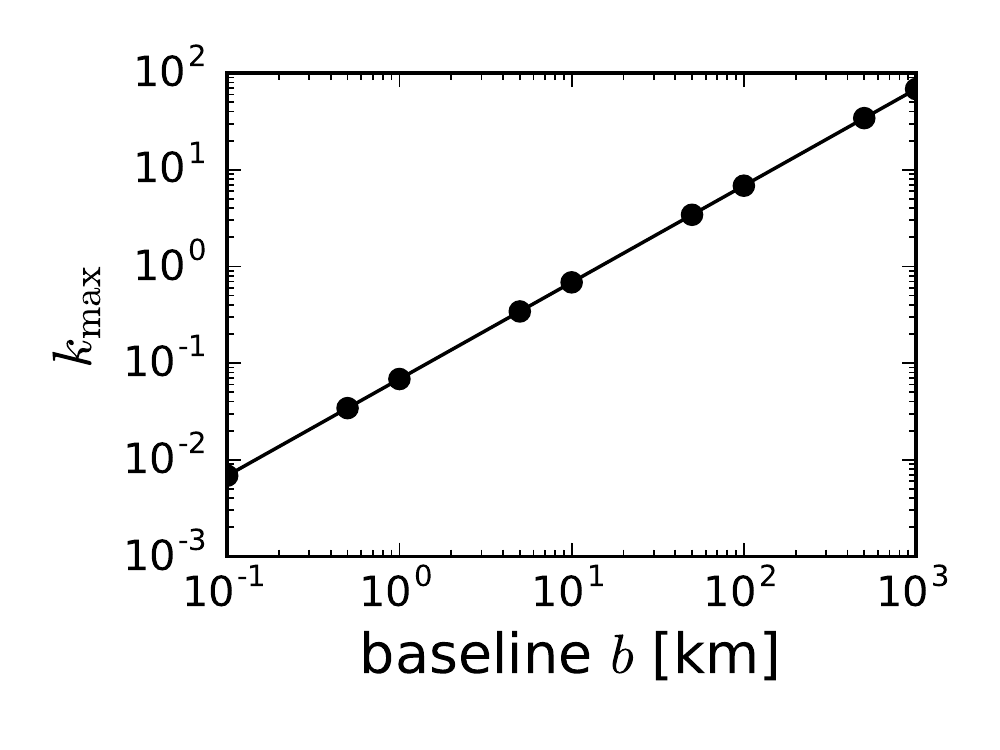}
}
\caption{Angular resolution $k_{\rm max}$ vs. array size for $z=65$, in the middle of our red-shift range $30<z<100$.}
\label{fig:baseline}
\end{figure}

It is very challenging to measure angular resolution with high precision. The maximum wavenumber observable is constrained by the baseline $b$ of the experiment, in km, as
\be
k_{\rm max} \simeq 2 \pi \nu_0 b \frac{1}{d(z) (1+z)} \frac{1}{c}
\ee
with $\nu_0$ the rest frequency of the 21-cm line in Hz, $c$ is the speed of light in km/s, and $d(z)$ is the conformal distance in Mpc given by the redshift window of the experiment. For reference we plot this function in Fig.~\ref{fig:baseline}. We also assume that the radial resolution matches the angular resolution, sharing the same $k_{\rm max}$. This is a conservative assumption as for current 21-cm experiments it is easier to improve the redshift resolution than the angular resolution. The minimum measurable mode $k_{\rm min}$ is limited by the radial distance accessible to an experiment with $30<z<100$ and we set it to $k_{\rm min}= 2\pi / (d_{z_{\rm max}} - d_{z_{\rm min}}) = 0.005$ Mpc$^{-1}$ in both the radial and angular direction.

Using these parameters, we plot the sensitivity as a function of array baseline in Fig. \ref{fig:fishermax} (right), assuming negligible instrumental noise. For a fixed primordial amplitude, the sensitivity of the oscillatory template does not depend on $\mu$ significantly. This is different for the non-oscillatory intermediate-bispectra template which has a mass-dependent power-law in its squeezed limit. This power-law is responsible for the larger signal to noise of the intermediate bispectra template, especially for larger $\nu$ (lower mass). As we explained above, for lower mass the template approaches the power-law of the local shape, which has most of its signal in the squeezed limit, the region we select with our shape envelope.

\begin{figure*}[t!]
\resizebox{0.8\hsize}{!}{
\includegraphics{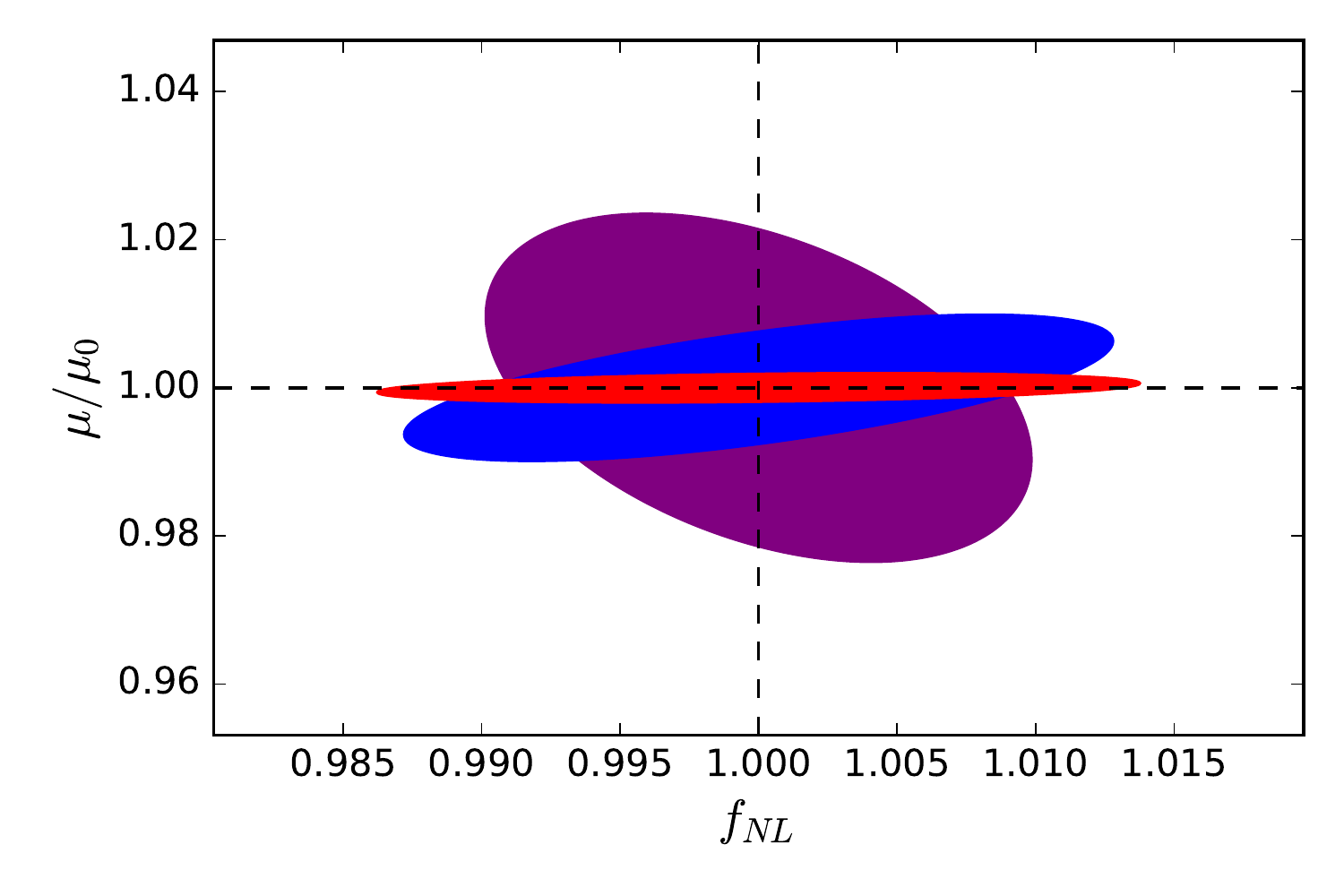}\includegraphics{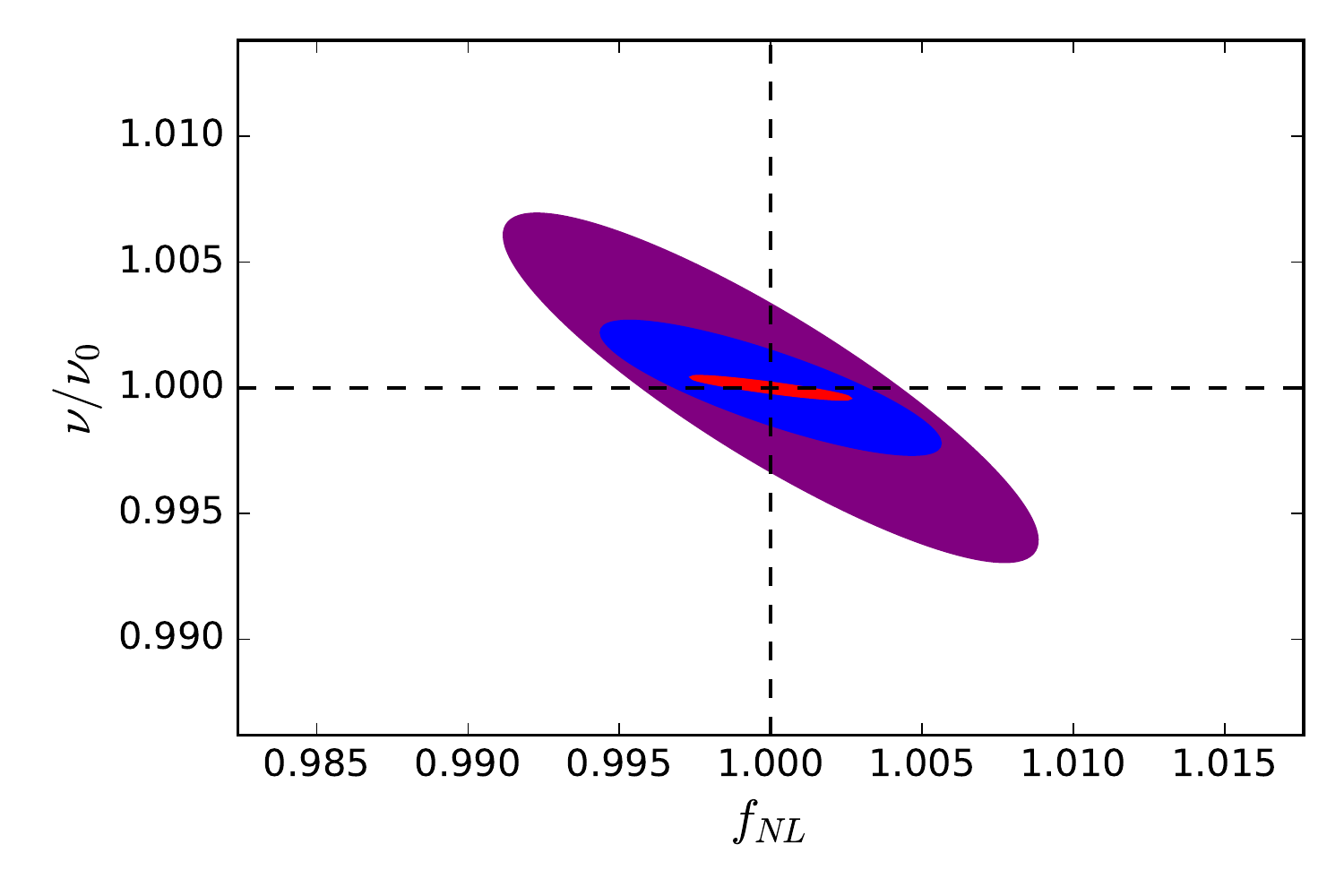}
}
\caption{Degeneracy of amplitude $f$ and mass parameter $\mu$ or $\nu$, for a cosmic variance limited experiment with $30<z<100$ and baseline $b=100$ km, assuming $f_{\rm NL}=1.0$. Left: Template Eq. \eqref{eq:template3}, with $\mu=0.7$ (purple), $\mu=1.0$ (blue), $\mu=3.0$ (red). The plot illustrates the improved mass measurement at higher mass. Right: Template Eq. \eqref{eq:template4}, with $\nu=0.8$ (purple), $\nu=1.0$ (blue), $\nu=1.4$ (red). }
\label{fig:fisherdeg}
\end{figure*}

We also examine the degeneracy of the amplitude and frequency measurements in Fig.~\ref{fig:fisherdeg} for an experiment with a baseline of $b=100$ km.  For the oscillatory template (left) the plot illustrates that larger masses $\mu$ allow a better mass determination, as more oscillations are visible. The correlation between amplitude and frequency also decreases for larger $\mu$, because more oscillations can be observed. Again for large $\nu$ the intermediate-bispectra template has a larger signal to noise for a given $f_{NL}$, so it allows for a more precise mass determination. Of course, non-Gaussianities with large amplitudes $f_{\rm NL}$ would allow a better determination of $\mu$ or $\nu$, perhaps allowing us to detect several masses independently. The relative uncertainty on the frequency scales with $f_{\rm NL}^{-1}$, as one can see from the Fisher matrix, so that for example $\Delta \mu/\mu \sim 1\% \times f_{\rm NL}^{-1}$ for $\mu=1.0$ in Fig.~\ref{fig:fisherdeg} (left). This property can be used to obtain sensitivities for other fiducial $f_{NL}$ values.

We compare the sensitivity for our template to that of more familiar shapes in Tab.~\ref{tab:fnl}, for a benchmark experiment with $30<z<100$ and $k_{\rm max}=10$ Mpc$^{-1}$. The local, equilateral and constant shape functions are normalized so that $S(1,1,1)=6 f_{\rm NL}$. We show values with and without the envelope function, to show how much of the phase space is cut off. We do not marginalize over the frequency or secondary non-Gaussianities here, as our aim is to compare the signal strength only.

Notice that these results are significantly more optimistic than those in Ref.~\cite{21cmFNL}. Part of the reason is that we do not marginalize over secondary non-gaussianities.
More important is our use of the $k$-space Fisher matrix on Eq.~\eqref{eq:Fisher}, which is the most optimistic case, whereas Ref.~\cite{21cmFNL} used $\ell$-space estimators, which discards significant line-of-sight information. The two cases can be taken as the optimistic and pessimistic limit, respectively.

As expected, the envelope cuts a large part of the signal of the equilateral shape, and almost nothing of the squeezed local model. From the constant model one can see that the envelope reduces the sensitivity by a factor 4, i.e. only about one in 16 triangles survives the cut. The massive particle template for the oscillatory shape with $\mu=1$ only makes physical sense with the envelope, as the full physical shape resembles the equilateral shape with a small oscillatory addition in the squeezed limit which we wish to pick up. The sensitivity of the shapes with envelope is dominated by their scaling properties in the squeezed limit, which of course favors in particular the local shape.

\begin{table}[hbtp!]
	
    \begin{tabular}{|l|l|}
         \hline
             local  & $\sigma_f=0.0003$   \\ \hline
             local with envelope & $\sigma_f=0.0003$ \\ \hline
             equilateral & $\sigma_f=0.0016$ \\ \hline
             equilateral with envelope & $\sigma_f=0.026$ \\ \hline
             constant $S=6$ & $\sigma_f=0.0011$ \\ \hline
             constant $S=6$ with envelope & $\sigma_f=0.0045$ \\ \hline
             $\mu=1$ template with envelope & $\sigma_f=0.0055$\\ \hline
             $\nu=1$ template without envelope& $\sigma_f=0.0008$ \\ \hline
             $\nu=1$ template with envelope & $\sigma_f=0.0017$ \\
         \hline
    \end{tabular}
\caption{Sensitivity to the amplitude of different primordial shapes for an experiment with $30<z<100$ and $k_{\rm max}=10$ Mpc$^{-1}$, without marginalisation over secondary non-gaussianity or shape parameters.}
\label{tab:fnl}
\end{table}

\subsection{Implications for inflation models}

An important model-independent aspect of the cosmological collider is that the characteristic power-law/oscillatory behaviors in the squeezed limit of the bispectrum encode the particle spectrum model-independently. The amplitude of the bispectrum does not affect this property, but determines whether these behaviors are observable.
In the previous sections, with this property in mind, we have studied how well  future 21-cm experiments can be used to measure these behaviors, given an amplitude $f_{\rm NL}$ of the bispectrum.

In inflation models, the amplitude $f_{\rm NL}$ crucially depends on the types of couplings between the heavy fields and the inflaton, and its value spans a wide range of values \cite{Chen:2009zp,Baumann:2011nk,Arkani-Hamed:2015bza,Chen:2015lza,Lee:2016vti}.
There are three types of couplings that give qualitatively different sizes of amplitudes.
In this section, using our results we discuss what types of couplings can potentially lead to measurable particle spectrum.

All the $f_{\rm NL}$'s in this subsection refer to those of the bispectra that carry the signature of the heavy fields, namely the intermediate/clock signal bispectra. In inflation models, these bispectra can coexist with other more conventional bispectra such as the local and equilateral bispectra. Notice in Tab.~\ref{tab:fnl} that the $f_{\rm NL}$ of the intermediate/clock signal bispectra is generally harder to constrain than the conventional ones.

$\bullet$ {\em Gravitational coupling}
We first consider inflation models in which the heavy fields couple to the inflaton field only through gravity. Such couplings exist in all models, so this is the minimal case \cite{Chen:2012ye}. In this case, the heavy fields contribute to the bispectrum through loop diagrams because they have to appear in pairs in each vertex.
As a comment, we note that none of the bispectrum diagrams in Fig.~\ref{fig:feynmans1} can arise in the gravitational-coupling-only models.

The amplitudes of the bispectrum in this case are suppressed by at least factors of slow-roll parameters and possibly by additional factors of scalar curvature (ignoring the cutoff dependence). So the expected $f_{\rm NL}$ is very small, i.e. $f_{\rm NL} \ll 10^{-2}$.
Our forecast shows that there is only a slim chance the particle spectrum could be observed even in the most optimistic experiment. It is a daunting task for several reasons. Firstly, only in the case with the most optimistic $k_{\rm max}$, $\sigma_{f_{\rm NL}}$ may be reached at the order $\CO(10^{-3})$ or $\CO(10^{-4})$. Such experiments would require a very long baseline (thousands of kilometers) filled array, which, because of ionispheric limitations, needs to be in space or on the moon. Secondly, the non-Gaussianity floor, $f_{\rm NL} \sim \CO(10^{-2})$, of the slow-roll inflation models is very likely reached \cite{Maldacena:2002vr} and becomes a background for the cosmological collider physics. This, however, may not be a problem for the oscillatory type of bispectra, similar to the analyses we did for the secondary non-Gaussianity in Sec.~\ref{Sec:Secondary}. Lastly, in this case theoretical predictions from loop diagrams need to be better understood.

$\bullet$ {\em Direct couplings.} The above minimal case geometrically means that the inflaton trajectory is strictly straight in the field space landscape, and realistically we expect some level of bending of the trajectory that naturally introduces direct couplings between the inflaton and heavy fields \cite{Chen:2009zp,Baumann:2011nk,Arkani-Hamed:2015bza,Lee:2016vti}.
There can also be other types of direct couplings that may not be easily visualized geometrically \cite{Baumann:2011nk}.
The direct couplings introduce diagrams such as A and C in Fig.~~\ref{fig:feynmans1}. They can potentially enhance the amplitude of the bispectrum up to $f_{\rm NL} < \CO(1)$. This theoretical constraint comes from requiring perturbative stability in the computation.

Our results indicate that the cases with $f_{\rm NL}\gtrsim 10^{-2}$ are potentially observable if one were to build an array with a baseline of order $\mathcal{O}(100)$ km, targeting $k_{\rm max}=10$ Mpc$^{-1}$. For example, in the case proposed in Ref.~\cite{Arkani-Hamed:2015bza}, $f_{\rm NL} \sim \epsilon \mpl^2 \lambda^2$ where $\lambda \gtrsim 1/\mpl$ parameterizes the size of the cubic coupling, and $\epsilon \lesssim 10^{-2}$. In the example in the Appendix A of \cite{Chen:2009zp}, several three-point diagrams give rise to $f_{\rm NL} \sim (\dot\theta_0/H)^2$ where $(\dot\theta_0/H)^2<1$ parameterizes the bilinear coupling. So in both cases, in principle some of these bispectra are observable at a level above the non-Gaussianity floor, which makes them interesting targets for high-energy physics studies.

$\bullet$ {\em Self-interactions.} Because heavy fields do not respect any slow-roll conditions, their self-interactions are potential sources of large non-linearities \cite{Chen:2009zp,Baumann:2011nk,Lee:2016vti}. This introduces diagrams such as B in Fig.~~\ref{fig:feynmans1}, which can boost the amplitude of the bispectrum to $f_{\rm NL} \gtrsim \CO(1)$. This is the most prominent case and may be accessible for experiments with baselines in the range of $\mathcal{O}(10)$ km, with angular resolution around $k_{\rm max}=1$ Mpc$^{-1}$.

This is also the case where the particle spectrum can be measured most precisely. If we were to find both large self interactions with a large-enough experiment, we could indeed make precise mass measurements, similar to those of a particle accelerator. Denoting $\mu=\sqrt{m^2/H^2-9/4}$, we forecasted that a 100-km baseline noiseless interferometer, observing at the dark ages, will be able to constrain $\Delta \mu/\mu \sim 1\% \times f_{\rm NL}^{-1}$ for $\mu=1.0$, and even better accuracy for higher $\mu$.

$\bullet$ Note that the discussion above applies to the case where the particle mass $\lesssim \CO(H)$. If particles have masses much larger than $H$, the amplitude of the corresponding bispectrum is suppressed by a Boltzmann factor \cite{Arkani-Hamed:2015bza}. (The details of the Boltzmann factor may change \cite{Lee:2016vti}.) This applies to all the interaction types mentioned above. So in a scale-invariant theory, the main interest of cosmological collider is fields with masses $\lesssim \CO(H)$, which is the focus of this paper. On the other hand, if inflation models contain features that break the scale-invariance, fields with $m > \CO(H)$ can be excited \cite{Chen:2011zf,Chen:2014cwa}. In this case, the massive fields can also imprint distinctive signatures in density perturbations through their classical oscillations, and the mass of the field that gets excited depends on the type of feature in the model and its value can be quite arbitrary, e.g., several orders of magnitude larger than $\CO(H)$.
Future experimental sensitivities on this type of signals are studied in \cite{Chen:2016zuu,Chen:2016vvw} using power-spectrum forecasts.

\section{Conclusions}\label{sec:conclusions}

Additional degrees of freedom, with masses comparable to the scale of inflation, are a generic prediction of a large class of inflationary models. In this paper we have examined the possibility of detecting such massive particles with future 21-cm probes. There is a large number of uncertain parameters in the problem, both of theoretical and experimental nature.
We have first constructed a template that should cover a large space of models,
enabling us to easily find the non-Gaussianity signals due to heavy fields during inflation, avoiding costly numerical integrals for every step.
We forecasted how well this template can be measured, albeit limiting our analysis to the optimistic cosmic-variance-limited case. In principle, if one could build a sufficiently large and sensitive instrument, foregrounds could be isolated due to their smooth frequency spectrum. Gravitational non-linearities during the dark ages can in principle be determined to exquisite precision and the computation of the brightness temperature is not limited by complicated astrophysics. While it might take decades to realize such an experiment, our analysis shows the benefits are impressive. Detecting particles with the cosmological collider would provide much more detailed information about the early Universe than non-Gaussianity from inflaton self-interactions alone. Our findings indicate that the minimal case, in which heavy fields only interact with the inflaton gravitationally, is unlikely to ever be detected (modulo some issues in theoretical understanding in this case). However, fields with large direct interactions with inflaton have a realistic chance to be detected if one can fully exploit the potential of 21-cm cosmology. Moreover, non-Gaussianity from self-interactions of the heavy fields themselves is even more promising, and would be a realistic target for experiments with a baseline of order $10$ km. In principle, the SKA will probe frequencies that could constrain the 21-cm field from redshifts as high as $z = 27$ and from earth, and redshifts up to $z \sim 45$ are possible. While many details remain to be worked out in the future, we have shown that cosmological collider physics may some day become a reality and shed light into the particle spectrum at the highest energies we might ever reach.

\section*{Acknowledgments}
We thank Francois Bouchet, Marc Kamionkowski, Benjamin Wandelt, Yi Wang, Guilherme Pimentel, Daniel Baumann, Hayden Lee and Zhong-Zhi Xianyu for helpful discussions. This research was supported in part by Perimeter Institute for Theoretical Physics. Research at Perimeter Institute is supported by the Government of Canada through the Department of Innovation, Science and Economic Development Canada and by the Province of Ontario through the Ministry of Research, Innovation and Science.
JBM is supported at JHU by the Simons Foundation.
XC is supported in part by National Science Foundation grant PHY-1066293.

\vspace{1cm}

\bibliography{COSMO}

\clearpage
\appendix

\section{Evaluation of the primordial bispectrum} \label{sec:analyticalshape}
\label{sec:primordialcalc}

\subsection{Short review of the in-in calculation}

The curvature bispectrum from inflation can be computed in the in-in formalism, and has been calculated for the shape we consider here in \cite{Chen:2015lza, Arkani-Hamed:2015bza}. We briefly review this calculation which serves as a starting point to our numerical calculations, and motivates the shape of the template. The key point is that the squeezed-limit shape that contains the information of heavy particle is independent of the details of the vertices.

The second order term of the in-in formula, which we need to evaluate the process in diagram A of Fig.~\ref{fig:feynmans1} is given by
\begin{widetext}
\bal
\langle \zeta^3 \rangle'
&\supset
\int_{t_0}^t d\tilde t_1 \int_{t_0}^t dt_1
\langle 0| H_I(\tilde t_1) \zeta_I^3 H_I(t_1) |0\rangle'
-2 {\rm Re} \left[ \int_{t_0}^t dt_1 \int_{t_0}^{t_1} dt_2
\langle 0| \zeta_I^3 H_I(t_1) H_I(t_2) |0\rangle' \right] \\
&= I_{\rm{NTO}} + I_{\rm{TO}} + {\rm c.c.}+ {\rm 2~perm.} ~,
\nonumber
\eal
We need two vertices for this process, a cubic term and a quadratic term. As an example, Ref.~\cite{Chen:2015lza} chooses
$\CL_3 \sim c_3 a^3 \dot\zeta^2 \sigma$ and $\CL_2 \sim c_2 a^3 \dot\zeta \sigma$, where $\zeta$ is the inflaton and $\sigma$ is the massive field.
With these vertices, we obtain for the non-time ordered integral
\bal
I_{\rm{NTO}}  = 2 u_{k_3}^* u_{k_1} u_{k_2}|_{\tau=0}
\left( \int_{-\infty}^0 d\tau_1 c_3 a^2 v_{k_3}^* u'^{*}_{k_1} u'^{*}_{k_2} \right)
\left( \int_{-\infty}^0 d\tau_2 c_2 a^3 v_{k_3} u'_{k_3} \right),
\eal
while for the time-ordered integral
\bal
I_{\rm{TO}}  = -2u_{k_3} u_{k_1} u_{k_2}|_{\tau=0} \left[ \int_{-\infty}^0 d\tau_1 c_3 a^2 v_{k_3}^* u'^{*}_{k_1} u'^{*}_{k_2}
\int_{\tau_1}^{0} d\tau_2 c_2 a^3 v_{k_3} u'^*_{k_3} +
\int_{-\infty}^0 d\tau_1 c_3 a^2 v_{k_3} u'^{*}_{k_1} u'^{*}_{k_2}
\int_{-\infty}^{\tau_1} d\tau_2 c_2 a^3 v^*_{k_3} u'^*_{k_3} \right]
\eal
\end{widetext}
The mode functions are given by
\bal
u_k &= \frac{H}{2\sqrt{\epsilon}\mpl} \frac{1}{k^{3/2}} (1+ik\tau) e^{-ik\tau} ~,
\nonumber \\
v_k &= -ie^{-\frac{\pi}{2}\mu + i \frac{\pi}{4}} \frac{\sqrt{\pi}}{2} H
(-\tau)^{3/2} H^{(1)}_{i\mu}(-k\tau).
\label{mode_functions_inflation}
\eal

\subsection{Numerical evaluation}

After inserting the mode functions, we change variables to $x=-k_{3}
\tau_1$ and $y=-k_3 \tau_2$  to make the integral depend only on a
single momentum variable $\alpha = k_{12}/k_3$. For the non time-ordered integral, using $a=-\frac{1}{\tau H}$, we find
\begin{widetext}
\be
I_{\rm{NTO}} = C \, (k_1 k_2 k_3)^{-1} \,k_3^{-3}
\left[ \int_{0}^\infty dx \, x^{(3/2)}  e^{-i \alpha x} H^{(2)}_{i\mu}(x) \right] \left[ \int_{0}^\infty  dy \, y^{(-1/2)} e^{i y} H^{(1)}_{i\mu}(y) \right],\label{eq:nto1}
\ee
where $C$ is a constant real positive number. We perform a Wick rotation as proposed for these integrals in \cite{Chen:2014cwa} to improve numerical convergence, obtaining
\be
I_{\rm{NTO}} = C \, (k_1 k_2 k_3)^{-1} \,k_3^{-3} (-i)
\left[ \int^{0}_{-\infty} dx \, x^{(3/2)}  e^{\alpha x} H^{(2)}_{i\mu}(i x) \right] \left[ \int_{0}^\infty  dy \, y^{(-1/2)} e^{-y} H^{(1)}_{i\mu}(i y) \right].
\label{eq:nto2}
\ee

With the same change of variables, and again performing a Wick rotation, the time-ordered integral is
\be
I_{\rm{TOpart2}} = C \, (k_1 k_2 k_3)^{-1} \,k_3^{-3} (-i) \int^{0}_{-\infty} dx \, x^{(3/2)} e^{\alpha x} H^{(1)}_{i\mu}(ix) \int^{x}_{-\infty} dy \, y^{(-1/2)}  e^{y} H^{(2)}_{i\mu}(i y), \label{eq:to2wick}\\
I_{\rm{TOpart1}} =  C \, (k_1 k_2 k_3)^{-1} \,k_3^{-3}  i \int^{0}_{-\infty} dx \, x^{(3/2)}  e^{\alpha x} H^{(2)}_{i\mu}(i x) \int^{0}_x  dy \, y^{(-1/2)} e^{y} H^{(1)}_{i\mu}(i y). \label{eq:to1wick}
\ee

Using the above expressions, we can now efficiently evaluate the bispectrum numerically. We define the bispectrum shape function as usual, so that the bispectrum phase space factor is factored out, i.e.,
\be
\label{eq:chenshape}
\langle \zeta^3 \rangle'
=
\frac{C}{(k_1k_2k_3)^2}
\left[ \frac{k_1 k_2}{k_3^2} I\left(\frac{k_1+k_2}{k_3}\right) + \frac{k_1 k_3}{k_2^2} I\left(\frac{k_1+k_3}{k_2}\right) + \frac{k_2 k_3}{k_1^2} I\left(\frac{k_2+k_3}{k_1}\right) \right]~.
\ee
The shape is therefore scale invariant under $k \rightarrow \lambda k$, so that the oscillations are a function of shape, not of scale. Results for the primordial shape function are shown in Fig.~\ref{fig:bispec1} where we show the characteristic logarithmic oscillation of the squeezed limit bispectrum. Collecting all factors, the dimensionless amplitude is given by
\be
C  = \frac{2\pi}{2^8} \left(\frac{H^4}{M_p^4} \frac{1}{\epsilon^2}\right) \frac{H^{-1}}{M_p^2 \epsilon} c_2 c_3 = 2 \pi A_s^2 \frac{1}{H M_p^2 \epsilon} c_2 c_3.
\ee
\end{widetext}

\subsection{The coupling of AHM}

The authors of Ref. \cite{Arkani-Hamed:2015bza} (AHM) used a slightly different coupling with more complicated momentum dependence. The $\mathcal{L}_3$ vertex uses the full spacetime derivative, i.e.,
\be
(\nabla \zeta)^2 \sigma \rightarrow \left[-(\partial_{\eta} \zeta)^2 + (\partial_x \zeta)^2\right] \sigma.
\ee
This term actually includes the previous vertex (containing the conformal time derivative). The $\mathcal{L}_2$ vertex is identical to the case of the previous section.

With these couplings, the diagram gives the contribution
\begin{widetext}
\be
I_{\rm{TO}}  &=& -2u_{k_3} u_{k_1} u_{k_2}|_{\tau=0} \left[ \int_{-\infty}^0 d\tau_1 c_3 a^2 v_{k_3}^* \left(-u'^{*}_{k_1} u'^{*}_{k_2} - \vec{k}_1 \cdot \vec{k}_2  u^{*}_{k_1} u^{*}_{k_2}  \right)
\int_{\tau_1}^{0} d\tau_2 c_2 a^3 v_{k_3} u'^*_{k_3} + \right. \nonumber \\
&& \left. \int_{-\infty}^0 d\tau_1 c_3 a^2 v_{k_3} \left(-u'^{*}_{k_1} u'^{*}_{k_2} - \vec{k}_1 \cdot \vec{k}_2  u^{*}_{k_1} u^{*}_{k_2}  \right)
\int_{-\infty}^{\tau_1} d\tau_2 c_2 a^3 v^*_{k_3} u'^*_{k_3} \right],
\ee
where we applied $\tilde{\nabla} = \{-\partial_{\tau}, i \vec{k}\}$. The second vertex is computed with respect to the background field. There is no spatial derivative and all that remains is the time derivative and a coupling that is a function of the background field. Using the mode functions we then obtain
\be
I_{\rm{TO}}  &=& \tilde{C} \frac{1}{k_1^3 k_2^3 k_3^3} \left[ \int_{-\infty}^0 \frac{d\tau_1}{\tau_1^2} \left(-k_1^2 k_2^2 \tau_1^2 -\vec{k_1} \cdot \vec{k_2} (1- i k_1 \tau_1)(1-ik_2\tau_1)\right)e^{i k_{12} \tau_1} v_{k_3}^*\int_{\tau_1}^{0} \frac{d\tau_2}{\tau_2^3}  k_3^2\tau_2 e^{i k_3 \tau_2} v_{k_3}+ \right. \nonumber \\
&& \left. \int_{-\infty}^0 \frac{d\tau_1}{\tau_1^2} \left(-k_1^2 k_2^2 \tau_1^2 -\vec{k_1} \cdot \vec{k_2} (1- i k_1 \tau_1)(1-ik_2\tau_1)\right)e^{i k_{12} \tau_1} v_{k_3}\int_{-\infty}^{\tau_1} \frac{d\tau_2}{\tau_2^3} k_3^2\tau_2 e^{i k_3 \tau_2} v_{k_3}^* \right],
\ee
with $k_{12} = k_1 + k_2$.
It can be shown \cite{Arkani-Hamed:2015bza} that the momentum dependence in the integral can be obtained by defining the operator
\be
\mathcal{O}_{12} =-\frac{1}{2} k_1 k_2 \left(k_3^2-k_{12}^2\right) \partial_{k_{12}} +\frac{1}{2} (k_1^2 + k_2^2 -k_3^2) (1-k_{12}\partial_{k_{12}}).
\ee
We then rewrite the integral as before in terms of $x$ and $y$ and the operator
\be
I_{\rm{TO}}  &=& \tilde{C} \frac{k_3^2}{k_1^3 k_2^3 k_3^3} \mathcal{O}_{12} \left[ \int_{-\infty}^0 \frac{d\tau_1}{\tau_1^2} e^{i k_{12} \tau_1} v_{k_3}^*\int_{\tau_1}^{0} \frac{d\tau_2}{\tau_2^2} e^{i k_3 \tau_2} v_{k_3}+  \int_{-\infty}^0 \frac{d\tau_1}{\tau_1^2} e^{i k_{12} \tau_1} v_{k_3}\int_{-\infty}^{\tau_1} \frac{d\tau_2}{\tau_2^2} e^{i k_3 \tau_2} v_{k_3}^* \right],
\ee
which simplifies to
\be
I_{\rm{TO}}  = \tilde{C}' \frac{k_3}{k_1^3 k_2^3 k_3^3} \mathcal{O}_{12} &\Big[& \int_{0}^{\infty} \frac{d x}{x^{1/2}} e^{-i \alpha x}  H^{(2)}_{i\mu}(x) \int_{x}^{0} \frac{dy}{y^{1/2}}  e^{-iy}  H^{(1)}_{i\mu}(y) \nonumber \\
&+&  \int_{0}^{\infty} \frac{dx}{x^{1/2}} e^{-i \alpha } H^{(1)}_{i\mu}(x)\int_{x}^{\infty} \frac{dy}{y^{1/2}} e^{-i y} H^{(2)}_{i\mu}(y) \Big],
\ee
\end{widetext}
which has the correct dimensions $k^{-6}$ as $\left[ \mathcal{O}_{12} \right] = k^2$. The integral is therefore identical to shape in Ref.~\cite{Chen:2015lza} discussed in the previous section, modulo different powers in $x$ and the application of the differential operator. We can integrate it in the same way as before after a Wick rotation. This also holds true for the non time-ordered integral $I_{\rm{NTO}}$. We plot the shape functions in Fig.~\ref{fig:bispec1}.

Different from the case of the previous section, AHM do not consider independent coupling strengths for $\mathcal{L}_2$ and $\mathcal{L}_3$. After introducing a direct coupling of form $\lambda (\nabla \phi)^2 \sigma$ between the inflaton $\phi$ and the massive field $\sigma$, the bilinear coupling is obtained by setting one leg of the vertex to the rolling inflaton background value $\dot{\phi}_0$.
Such a procedure can be naturally realized in terms of a turning trajectory.
From an EFT point of view, one expects $\lambda$ to be of order $\lambda \gtrsim \frac{1}{M_{PL}}$. Going back to curvature perturbations $\zeta$ one finds the slow roll suppressed amplitude of AHM
\be
\tilde{C} = 4 A_s^2 \epsilon M_p^2 \lambda^2
\ee


\section{Secondary shapes}
\label{app:secondaries}

The explicit form of the 21 secondary shapes of non-Gaussianity
was calculated in Ref.~\cite{21cmFNL}. Here we will review the different
families of secondary non gaussianities, and how they arise.

For convenience, let us name $P_{i}$ is the matter power spectrum of $k_i$, so
\be
P_i \equiv P_\zeta(k_i) \mathcal T^2(k_i),
\ee
with $\mathcal T$ the matter transfer function. We also define $\mu_i$ as the
line-of-sight angle of the $i$th wavenumber, i.e.
\be
\mu_i = k_{i,||}/k_i,
\ee
which ranges from $-1$ to 1.

The 21-cm temperature $T_{21}$ depends on the baryon perturbation $\delta_b$,
and the velocity perturbation $\delta_v$, non linearly, which creates
non-Gaussianities in $T_{21}$ even if the matter was distributed Gaussian.
From this non linearity we find bispectra of the type
\begin{align}
\left <\delta T(\bsk_1) \delta T(\bsk_2) \delta T(\bsk_3)\right> &\supset \left <\delta_b(\bsk_1) \delta_b(\bsk_2) \left(\delta_b \delta_b\right)(\bsk_3)\right> \nonumber \\
&= 2 P_1 P_2,
\end{align}
and similar combinations with $\delta_v$, which introduce $\mu_i$ factors.

Additionally, matter is not distributed gaussianly, due to the non-linear evolution of the density perturbations. In general, denoting $\delta_b^{(2)}$ as the second-order baryon perturbation, it can be shown that
\be
\delta_b^{(2)}(\bsk) = \int d^3\bsk' F(\bsk,\bsk')\delta_b^{(1)}(\bsk'-\bsk)\delta_b^{(1)}(\bsk'),
\ee
where $\delta_b^{(1)}$ is the linear-order perturbation, and $F$ is a kernel, that can be written as~\cite{Bernardeau:2001qr}
\be
F(\bsk,\bsk') = c_1 + c_2 \,\bsk \cdot \bsk' \left( \!\dfrac{1}{k'^2}+\dfrac{1}{k^2}\!\right) + c_3 \dfrac {\left(\bsk \cdot \bsk'\right)^2} {k^2k'^2},\ee
where the $c_i$ parameters can be computed for a given cosmology, and for an Einstein-de-Sitter Universe are $(c_1,c_2,c_3)=(5/7,1/2,2/7)$. Given the precision required when subtracting this shape, which would be orders of magnitude above the primordial shape, we marginalize over each of the $c_i$ independently. This creates non-Gaussianities of the type
\begin{align}
\left <\delta T(\bsk_1) \delta T(\bsk_2) \delta T(\bsk_3)\right> &\supset \left < \delta_b^{(1)}(\bsk_1)\delta_b^{(1)}(\bsk_2)\delta_b^{(2)}(\bsk_3) \right> \nonumber \\
&= 2 F(\bsk_1,\bsk_2) P_1 P_2,
\end{align}
and similar with combinations containing $\delta_v$, although with a kernel including different coefficients $c_i$.

We can then write the 21 components as
	\begin{align}
		B_{\rm sec,1}&=\left (\dfrac{k_1}{k_2}+\dfrac{k_2}{k_1} \right) \left (\dfrac{-k_1^2-k_2^2+k_3^2}{2k_1 k_2} \right) P_1 P_2 \\ \nonumber
		B_{\rm sec,2}&=\left( \dfrac{-k_1^2-k_2^2+k_3^2}{2k_1 k_2} \right)^2  P_1 P_2 \\ \nonumber
		B_{\rm sec,3}&=P_1 P_2(\mu_1^2+\mu_2^2) \\ \nonumber
		B_{\rm sec,4}&=\left (\dfrac{k_1}{k_2}+\dfrac{k_2}{k_1} \right) \left (\dfrac{-k_1^2-k_2^2+k_3^2}{2k_1 k_2} \right) P_1 P_2 (\mu_1^2+\mu_2^2) \\ \nonumber
		B_{\rm sec,5}&=\left(\dfrac{-k_1^2-k_2^2+k_3^2}{2k_1 k_2} \right)^2 P_1 P_2 (\mu_1^2+\mu_2^2) \\ \nonumber
		B_{\rm sec,6}&=P_1 P_2\mu_1^2\mu_2^2 \\ \nonumber
		B_{\rm sec,7}&=\left (\dfrac{k_1}{k_2}+\dfrac{k_2}{k_1} \right) \left (\dfrac{-k_1^2-k_2^2+k_3^2}{2k_1 k_2} \right) P_1 P_2  \mu_1^2\mu_2^2 \\ \nonumber
		B_{\rm sec,8}&=\left(  \dfrac{-k_1^2-k_2^2+k_3^2}{2k_1 k_2} \right)^2  P_1 P_2  \mu_1^2\mu_2^2 \\ \nonumber
		B_{\rm sec,9}&=P_1 P_2 \, \mu_1^2\mu_2^2\mu_3^2 \\ \nonumber
		B_{\rm sec,10}&=\left (\dfrac{k_1}{k_2}+\dfrac{k_2}{k_1} \right) \left (\dfrac{-k_1^2-k_2^2+k_3^2}{2k_1 k_2} \right) P_1 P_2 , \mu_1^2\mu_2^2\mu_3^2 \\ \nonumber
		B_{\rm sec,11}&=\left (\dfrac{-k_1^2-k_2^2+k_3^2}{2k_1 k_2}\right)^2  P_1 P_2 , \mu_1^2\mu_2^2\mu_3^2 \\ \nonumber
		B_{\rm sec,12}&=P_1 P_2\, \mu_3^2 \\ \nonumber
		B_{\rm sec,13}&=\left (\dfrac{k_1}{k_2}+\dfrac{k_2}{k_1} \right) \left (\dfrac{-k_1^2-k_2^2+k_3^2}{2k_1 k_2} \right) P_1 P_2 \,\mu_3^2 \\ \nonumber
		B_{\rm sec,14}&=\left (\dfrac{-k_1^2-k_2^2+k_3^2}{2k_1 k_2}\right)^2  P_1 P_2(\mu_3^2) \\ \nonumber
		B_{\rm sec,15}&=P_1 P_2(\mu_1^2+\mu_2^2) \\ \nonumber
		B_{\rm sec,16}&=\left (\dfrac{k_1}{k_2}+\dfrac{k_2}{k_1} \right) \left (\dfrac{-k_1^2-k_2^2+k_3^2}{2k_1 k_2} \right)  P_1 P_2(\mu_1^2+\mu_2^2) \mu_3^2 \\ \nonumber
		B_{\rm sec,17}&=\left(  \dfrac{-k_1^2-k_2^2+k_3^2}{2k_1 k_2} \right)^2  P_1 P_2\, (\mu_1^2+\mu_2^2) \mu_3^2 \\ \nonumber
		B_{\rm sec,18}&=P_1 P_2(\mu_1^2+\mu_2^2) (\mu_1^2+\mu_2^2) \\ \nonumber
		B_{\rm sec,19}&=P_1 P_2(\mu_1^2+\mu_2^2) \\ \nonumber
		B_{\rm sec,20}&=P_1 P_2\, \mu_1^2\mu_1^2\mu_2^2\mu_2^2\nonumber\\
		B_{\rm sec,21}&=P_1 P_2,   \nonumber
	\end{align}			
where each component has two additional permutations,
obtained by $(1\rightarrow2$, $2 \rightarrow3$, $3\rightarrow1)$, and
$(1\rightarrow3$, $2 \rightarrow1$, $3\rightarrow2)$.

\section{Convention of $f_{\rm NL}$} \label{sec:bispectrumNorm}

In this paper we defined the power spectrum as in the Planck inflation paper~\cite{PlanckInflation2013} eq. (25) and eq. (10) to be
\be
\langle \zeta^2 \rangle \equiv (2 \pi)^3 \delta_D(\mathbf{k}_{12}) \frac{2 \pi^2 A_s}{k^3}
\ee
where $A_s \approx 2.2\times 10^{-9}$ at $k_*=0.05{\rm Mpc}^{-1}$ (Table 3), and we neglect the tilt $n_s$. We also defined the bispectrum as \cite{Chen:2010xka,Wang:2013eqj}
\be
\langle \zeta^3 \rangle \equiv (2\pi)^3 \delta_D(\mathbf{k}_{123}) \frac{A^2}{(k_1k_2k_3)^2} S(k_1,k_2,k_3).
\label{Our_bispectrum}
\ee
where $A = 2 \pi^2 A_s$. The Planck non-Gaussianity paper~\cite{PlanckNGs2013} works with the potential $\Phi$ related by $\zeta=(5/3) \Phi$ and defines $f_{NL}$ by
\bea
\langle \Phi^3 \rangle  \equiv (2\pi)^3 \delta_D(\mathbf{k}_{123}) \frac{A_\Phi^2}{(k_1k_2k_3)^2} S_\Phi(k_1,k_2,k_3).
\label{Planck_bispectrum}
\eea
where $A_\Phi$ is defined by
\bea
\langle \Phi^2 \rangle = (2\pi)^3 \delta^3(\bk_1+\bk_2) \frac{A_\Phi}{k_1^3} ~,
\eea
We thus see that there is a factor of $(5/3)$ between the two definitions of $S$. This order one factor can be taken care of by a detailed definition of $f_{\rm NL}$. For example for the local non-Gaussianity,
in (\ref{Our_bispectrum}), Ref.~\cite{Maldacena:2002vr} uses
\bea
S = 4 \times \frac{3}{10} f_{\rm NL} \left( \frac{k_1^2}{k_2k_3} + {\rm 2~ perm} \right)
\eea
In (\ref{Planck_bispectrum}), Planck uses
\bea
S_\Phi =  2 f_{\rm NL} \left( \frac{k_1^2}{k_2k_3} + {\rm 2~ perm} \right)
\eea
So the extra $5/3$ is absorbed in the definition of $f_{\rm NL}$.

\end{document}